\documentclass[11pt,aps,prd,nofootinbib,superscriptaddress,floatfix,preprint]{revtex4}
\usepackage[T1]{fontenc}
\usepackage[latin9]{inputenc}
\usepackage{textcomp, amsmath, amssymb, cancel, graphicx, amsthm, color, fancyhdr, slashed, mathrsfs, tikzsymbols, float, feynmf, dsfont, xcolor, braket, gensymb, latexsym, tikz, adjustbox, lipsum, rotating, wrapfig}
\usepackage[unicode=true, bookmarks=false, breaklinks=false,pdfborder={0 0 1},backref=section,colorlinks=false]{hyperref}
\hypersetup{colorlinks,bookmarksopen,bookmarksnumbered,linkcolor=blu,pdfstartview=FitH,urlcolor=rossos,citecolor=rossos}

\usepackage[justification=raggedright, margin=5mm,font=sf,small,labelfont=bf, format=plain]{caption}
\definecolor{rosso}{cmyk}{0,1,1,0.4}
\definecolor{rossos}{cmyk}{0,1,1,0.55}
\definecolor{rossoc}{cmyk}{0,1,1,0.2}
\definecolor{blu}{cmyk}{1,1,0,0.3}
\definecolor{blus}{cmyk}{1,1,0,0.6}
\definecolor{bluc}{cmyk}{1,1,0,0.1}
\definecolor{verde}{cmyk}{0.92,0,0.59,0.25}
\definecolor{verdec}{cmyk}{0.92,0,0.59,0.15}
\definecolor{verdes}{cmyk}{0.92,0,0.59,0.4}

\definecolor{lime}{HTML}{A6CE39}
\DeclareRobustCommand{\orcidicon}{
 \begin{tikzpicture}
 \draw[lime, fill=lime] (0,0) circle [radius=0.2];
 \node[white] at (0,0) {{\fontfamily{qag}\selectfont \tiny ID}};
 \draw[white, fill=white] (-0.0625,0.095) circle [radius=0.007];
 \end{tikzpicture}
 \hspace{-2mm} }

\foreach \x in {A, ..., Z}{\expandafter\xdef\csname orcid\x\endcsname{\noexpand\href{https://orcid.org/\csname orcidauthor\x\endcsname}
			{\noexpand\orcidicon}} }

\newcommand{\arXivold}[2]{\href{http://arxiv.org/pdf/#1}{{\tt #2/#1}}}

\begin{document}
\title{\color{bluc}Novel and Updated Bounds on Flavor-violating Z Interactions in the Lepton Sector}

\author{Fayez Abu-Ajamieh\orcidA{}}
\email{fayezajamieh@iisc.ac.in}
\affiliation{Center for High Energy Physics, Indian Institute of Science, Bangalore, India}

\author{Amine Ahriche\orcidB{}}
\email{ahriche@sharjah.ac.ae}
\affiliation{Department of Applied Physics and Astronomy, University of Sharjah,
P.O. Box 27272 Sharjah, UAE}
\affiliation{Research Institute of Sciences and Engineering, University of Sharjah,
P.O. Box 27272 Sharjah, UAE}

\author{Nobuchika Okada\orcidC{}}
\email{okadan@ua.edu}
\affiliation{ Department of Physics and Astronomy, University of Alabama,\\
 Tuscaloosa, Alabama 35487, USA}

\begin{abstract}
We investigate the experimental bounds on the Flavor-Violating (FV)
couplings of the $Z$ boson to the charged leptons. In addition to
the direct LHC searches for FV $Z$ decays to leptons, we investigate
indirect bounds from flavor-conserving $Z$ decays to leptons at 1-loop,
bounds from LEP searches, Electroweak Precision Observables (EWPO),
$\ell_{i}\to\ell_{j}\gamma$ decays, $\ell_{i}\to 3\ell_{j}$ decays,
the muon lifetime, FV meson decays to leptons, FV $\tau$ decays to
$\mu(e)$ + mesons, muon conversion in nuclei, the Michel parameters,
and from muonium-antimuonium oscillations. For FV $Z$ couplings to
$\tau\mu$, we find that $\tau \to \mu \gamma$ yields the strongest
bounds, with a level reaching $\mathcal{O}(10^{-5})$, followed by
bounds from $\tau \to 3\mu~(\mu ee)$. For FV $Z$ couplings to $\tau e$,
we find that the strongest bounds arise from the decay $\tau\to \mu\mu e$,
reaching $\mathcal{O}(10^{-7})$ as well, with bounds from $\tau \to 3e$
also yielding strong bounds. For FV $Z$ couplings to $\mu e$, we
find that the strongest bounds are obtained from the decay $\mu\to 3e$,
reaching $\mathcal{O}(10^{-11})$, with bounds from $\mu\to e\gamma$,
muon conversion, and $K_{L}^{0}\to \mu e$ also providing strong
bounds. We also study projections from future experiments, such as
the FCC-ee, Belle II and Mu3e. For the $Z$ couplings to $\tau\mu$,
we find that future experiments could improve the bound to $\mathcal{O}(10^{-6})$,
whereas for the $Z$ couplings to $\tau e$, we find that future experiments
could improve the bound to $\mathcal{O}(10^{-8})$, and for the $Z$
couplings to $\mu e$, they could improve the bound to $\mathcal{O}(10^{-13})$.
\end{abstract}

\maketitle

\section{Introduction}
The discovery of the Higgs boson at the Large Hadron Collider (LHC)
marked a significant milestone in particle physics~\cite{ATLAS:2012yve},
completing the Standard Model (SM) framework. Despite this achievement,
the SM falls short of addressing several fundamental questions, such
as the origin of neutrino masses, the matter-antimatter asymmetry
in the Universe, the nature of dark matter and dark energy, and the
strong CP problem. These unresolved issues strongly suggest that the
SM is not the ultimate theory but rather an Effective Field Theory
(EFT) valid up to a certain Ultraviolet (UV) energy scale $\Lambda$.
At this scale, New Physics (NP) phenomena are expected to emerge,
manifesting as small deviations from the SM predictions that are suppressed
by powers of $\Lambda$~\cite{Chang:2019vez, Abu-Ajamieh:2020yqi, Abu-Ajamieh:2021egq, Abu-Ajamieh:2022ppp, 
Dawson:2022zbb, Abu-Ajamieh:2022dtm, Abu-Ajamieh:2021vnh}.
This perspective has motivated extensive efforts to explore Beyond
the Standard Model (BSM) physics through both theoretical models and
experimental searches.

There have been significant efforts to search for BSM physics, with
many proposals put forward. One such proposal involves searching for
Flavor Violation (FV). It is known that the SM lacks any Flavor-Changing
Neutral Currents (FCNC)~\cite{Glashow:1970gm} at tree level. The
QED photon couples universally to charged fermions, depending only
on their electric charge. In the SM Higgs sector, the Yukawa couplings
do not violate flavor, and the SM $Z$ boson also preserves flavor
at tree level. However, incorporating FCNC in BSM physics is relatively
straightforward with some minor assumptions.

FV in the Higgs sector has attracted significant attention over the
years (see for instance,~\cite{Bjorken:1977vt,McWilliams:1980kj,Shanker:1981mj,Barr:1990vd,Babu:1999me,Diaz-Cruz:1999sns,Han:2000jz,Blanke:2008zb,Casagrande:2008hr,Giudice:2008uua,Aguilar-Saavedra:2009ygx,Albrecht:2009xr,Buras:2009ka,Agashe:2009di,Goudelis:2011un,Arhrib:2012mg,McKeen:2012av,Azatov:2009na,Blankenburg:2012ex,Kanemura:2005hr,Davidson:2010xv,Harnik:2012pb,Abu-Ajamieh:2022nmt,Abu-Ajamieh:2023qvh,Abu-Ajamieh:2025jsz}).
As pointed out in~\cite{DAmbrosio:2002vsn}, it is well known that
in the SM, there exists an approximate $U(1)^{5}$ symmetry of light
quarks, which is broken by the Yukawa couplings. Thus, UV interactions
with the Higgs need not necessarily preserve flavor. However, the
fact that FV at $\Lambda\sim$ a few TeV is experimentally excluded
led the authors to propose the Minimal Flavor Violation (MFV) framework,
in which all flavor- and CP-violating interactions are linked to known
Yukawa interactions.

On the other hand, less attention has been given to studying FV in
the $Z$ sector~\cite{Brignole:2004ah,Davidson:2012wn,Goto:2015iha,Kamenik:2023hvi,Jueid:2023fgo}
(see also~\cite{Calibbi:2021pyh} where they provide a comprehensive
study of FV in the $Z$ sector based on the SMEFT) or the $Z'$ sector~\cite{Chivukula:2002ry,Erler:2009jh,Aranda:2010cy,Murakami:2001cs,Altmannshofer:2016brv}.
In this work, we aim to update the bounds on FV $Z$ interactions
with charged leptons and establish novel constraints that, to the
best of our knowledge, have not been previously calculated. We review
the bounds arising from direct searches at the LHC for $Z\to\ell_{i}\ell_{j}$
decays, FV loop corrections to the flavor-conserving $Z\to\ell_{i}\ell_{i}$
decays, LEP searches, electroweak precision observables (EWPO), lepton
flavor-violating decays such as $\ell_{i}\to3\ell_{j}$, $\ell_{i}\to\ell_{j}\gamma$
decays, muon conversion in nuclei, and muonium-antimuonium oscillation.
Additionally, we derive novel bounds from processes such as meson
decays, the muon lifetime, the Michel parameters, and from $\tau$
decays to $\mu(e)$ + mesons.

In general, we find that bounds extracted from indirect searches are
stronger than those from the LHC's direct searches by several orders
of magnitude. Specifically, for the FV $Z$ couplings to $\tau\mu$,
we find that $\tau\to\mu\gamma$ provides the strongest constraints,
reaching a level of $\mathcal{O}(10^{-5})$. Bounds from the decays
$\tau\to3\mu$ and $\tau\to\mu ee$ also impose strong constraints,
at the level of $\mathcal{O}(10^{-4})-\mathcal{O}(10^{-5})$, compared
to $\mathcal{O}(10^{-3})$ from direct searches. Additionally, we
find that bounds from FV leptonic meson decays can be as strong as
$\mathcal{O}(10^{-2})-\mathcal{O}(10^{-4})$, whereas bounds from
$\tau$ decays involving mesons could reach $\mathcal{O}(10^{-4})-\mathcal{O}(10^{-5})$.
Other constraints, such as those from the EWPO and flavor-conserving
$Z$ decays, provide less stringent bounds. Finally, we find that
future experiments on $\tau\to\mu\gamma$ could improve the bound
to $\mathcal{O}(10^{-6})$.

As for the bounds on the FV $Z$ couplings to $\tau e$, we find that
$\tau$ decays to $3e$ and $\mu\mu e$ provide the strongest constraints,
reaching $\mathcal{O}(10^{-7})$, compared to $\mathcal{O}(10^{-3})$
from direct searches. We also find that $\tau\to e\gamma$ can impose
strong limits of $\mathcal{O}(10^{-5})$. Bounds from meson decays
can reach $\mathcal{O}(10^{-2})-\mathcal{O}(10^{-4})$, whereas those
from $\tau$ decays involving mesons can impose constraints at the
level of $\mathcal{O}(10^{-4})-\mathcal{O}(10^{-5})$. Other constraints,
such as those from LEP searches and loop corrections to flavor-conserving
$Z$ decays to leptons, are much weaker and reach at best of $\mathcal{O}(10^{-2})$.
Future experiments on $\tau\to3e$ may improve the bound to $\mathcal{O}(10^{-8})$.

Finally, the strongest bounds on FV $Z$ couplings to $\mu e$ arise
from the $\mu\to3e$ decay, which reach $\mathcal{O}(10^{-11})$,
compared to only $\mathcal{O}(10^{-4})$ from direct searches. Constraints
from $\mu\to e$ conversion and $\mu\to e\gamma$ are also stringent,
reaching $\mathcal{O}(10^{-8})-\mathcal{O}(10^{-11})$. Bounds from
meson decays to $\mu e$ can be as strong as $\mathcal{O}(10^{-2})-\mathcal{O}(10^{-8})$,
whereas other constraints, such as those from LEP searches, flavor-conserving
$Z$ decays, muon lifetime, the Michel parameters, meson decays, and
muonium-antimuonium oscillations, yield weaker bounds of at most $\mathcal{O}(10^{-2})-\mathcal{O}(10^{-3})$.
Future experiments, particularly on $\mu\to3e$, could further improve
the bound to $\mathcal{O}(10^{-13})$. In general, we find that the
most promising experiments to search for FV in the $Z$ interaction
with leptons include the decays of $\ell_{i}$ to $3\ell_{j}$ and
$\ell_{j}\gamma$, in addition to muon conversion. Furthermore, we
find that both meson decays and $\tau$ decays involving mesons are
also promising channels to search for FV and deserve attention. One
has to mention that a similar analysis has been performed in the quark
sector, where different bounds have been established on FV Z-quarks
couplings~\cite{Abu-Ajamieh:2025jov}.

This paper is organized as follows: In Section~\ref{sec:2} we present
our framework for describing FV in the $Z$ interaction with fermions.
In Section~\ref{sec:3}, we discuss all the bounds and projections
in detail. We relegate some of the technical details to the appendices.
We finally present our conclusions and future directions in Section
\ref{sec:4}.

\section{The Framework}

\label{sec:2} As alluded to above, in the SM the $Z$ sector is flavor-conserving
at tree level. More specifically, the coupling of the $Z$ boson to
fermions can be written as 
\begin{equation}
\mathcal{L}_{Zff}^{\text{SM}}=-Z_{\mu}\bar{f}\gamma^{\mu}(g_{L}^{f}P_{L}+g_{R}^{f}P_{R})f,\label{eq:Sm_Z_coupling}
\end{equation}
where $P_{L,R}=\frac{1}{2}(1\mp\gamma^{5})$ and 
\[
g_{L}^{f}=\frac{g_{2}}{\cos\theta_{W}}(T_{fL}^{3}-Q_{f}\sin^{2}\theta_{W}),\hspace{5mm}g_{R}^{f}=\frac{-g_{2}}{\cos\theta_{W}}(Q_{f}\sin^{2}\theta_{W}),
\]
with $g_{2}$ being the $SU(2)_{L}$ gauge coupling, $T_{fL}^{3}$
being the $3^{\text{rd}}$ component of the weak isospin, $\theta_{W}$
being the weak mixing angle and with both $g_{L}^{f}$ and $g_{R}^{f}$
being real in the SM. Working in a bottom-up approach, it is fairly
easy to generalize Eq.~(\ref{eq:Sm_Z_coupling}) to include FV simply
by promoting $g_{L,R}^{f}$ to become non-diagonal matrices. As such,
the FV version of Eq.~(\ref{eq:Sm_Z_coupling}) becomes~\footnote{The fact that each of the Lagrangian terms (\ref{eq:Sm_Z_coupling})
and (\ref{eq:FV_Z_coupling}) is self-Hermitian can be seen as follows:
$\left[\bar{f}_{i}\gamma^{\mu}(g_{L}^{ij}P_{L}+g_{R}^{ij}P_{R})f_{j}\right]^{\dagger}=\bar{f}_{j}(g_{L}^{ij*}P_{R}+g_{R}^{ij*}P_{L})\gamma^{\mu}f_{i}=\bar{f}_{j}\gamma^{\mu}(g_{L}^{ij*}P_{L}+g_{R}^{ij*}P_{R})f_{i}=\bar{f}_{i}\gamma^{\mu}(g_{L}^{ji*}P_{L}+g_{R}^{ji*}P_{R})f_{j}$,
which means $g_{L}^{ij}=g_{L}^{ji*}$ and $g_{R}^{ij}=g_{R}^{ji*}$.} 
\begin{equation}
\mathcal{L}_{Zf_{i}f_{j}}^{\text{FV}}=-Z_{\mu}\bar{f}_{i}\gamma^{\mu}(g_{L}^{ij}P_{L}+g_{R}^{ij}P_{R})f_{j},\label{eq:FV_Z_coupling}
\end{equation}
with the diagonal elements of the matrices $g_{L,R}^{ij}$ corresponding
to the SM flavor-conserving couplings 
\begin{align}
g_{L}^{\text{SM}}=g_{L}^{ii}\equiv g_{L}^{f},\hspace{5mm}g_{R}^{\text{SM}}=g_{R}^{ii}\equiv g_{R}^{f}.\label{eq:SM_gLR}
\end{align}

In general, the coupling matrix with the elements $g_{L,R}^{ij}=g_{L,R}^{ji*}$
is complex Hermitian. The Lagrangian in Eq.~(\ref{eq:FV_Z_coupling})
is an EFT that arises from an unknown UV sector at some scale of NP
$\Lambda$. However, we can match the BSM contributions to $g_{L,R}^{ij}$
to higher-dimensional operators in the SMEFT. For example,~\cite{Davidson:2012wn}
studied FV $Z$ decays to $\tau\mu$ and $\mu e$ using the dimension-6
FV operators in the basis developed in~\cite{Buchmuller:1985jz}.
However, here we find it more convenient to use the so-called Warsaw
basis~\cite{Grzadkowski:2010es}. We point out that $Z$ decays within
the SMEFT were studied in~\cite{Dawson:2018jlg}, but with no treatment
of FV. At dimension-6, the following FV operators can contribute to
FV $Z$ interactions 
\begin{align}
\mathcal{O}_{\text{SMEFT}}^{6} & :(H^{\dagger}i\overleftrightarrow {D}_{\mu}H)(\bar{\ell}_{i}\gamma^{\mu}\ell_{j}),(H^{\dagger}i\overleftrightarrow {D^{I}}_{\mu}H)(\bar{\ell}_{i}\sigma^{I}\gamma^{\mu}\ell_{j}),(H^{\dagger}i\overleftrightarrow {D}_{\mu}H)(\bar{e}_{i}\gamma^{\mu}e_{j}),\nonumber\\
 & \hspace{20mm}(\bar{\ell}_{i}\sigma^{\mu\nu}e_{j})(\sigma^{I}HW_{\mu\nu}^{I}),(\bar{\ell}_{i}\sigma^{\mu\nu}e_{j})(HB_{\mu\nu}),
\end{align}
\label{eq:SMEFT-6}
where $i=j$ corresponds to flavor-conserving operators and $i\neq j$
to FV ones, $W_{\mu\nu}^{I}=\partial_{\mu}W_{\nu}^{I}-\partial_{\nu}W_{\mu}^{I}-g\epsilon^{IJK}W_{\mu}^{J}W_{\nu}^{K}$,
$B_{\mu\nu}=\partial_{\mu}B_{\nu}-\partial_{\nu}B_{\mu}$, $H^{\dagger}i\overleftarrow {D}_{\mu}H=iH^{\dagger}(D_{\mu}-\overleftarrow{D}_{\mu})H$,
$H^{\dagger}i\overleftarrow {D}_{\mu}^{I}H=iH^{\dagger}(\sigma^{I}D_{\mu}-\overleftarrow{D}_{\mu}\sigma^{I})H$,
and the covariant derivative is given by $D_{\mu}=\partial_{\mu}+ig_{1}YB_{\mu}+\frac{i}{2}g_{2}W_{\mu}^{I}\sigma^{I}$.
Notice here that since the first three operators can contribute to
the off-diagonal FV elements of $g_{L,R}^{ij}$, as well as the diagonal
SM ones, the contributions to the SM couplings must be suppressed
to satisfy the experimental constraints on $g_{L,R}$. Therefore,
the Lagrangian in Eq.~(\ref{eq:FV_Z_coupling}) involves some fine-tuning
to keep the diagonal contributions suppressed while allowing for significant
deviations in the off-diagonal ones. Such fine-tuning might appear
unnatural, however, we show below that it is quite possible within
the relevant FV dimension-6 SMEFT operators without radical assumptions.
Here, $g_{1}$ is the $U(1)_{Y}$ gauge coupling.

Notice that the operators in Eq.~(\ref{eq:SMEFT-6}) are in the unbroken
phase. It is more convenient to work in the broken phase where the
physical $Z$ boson is manifest. The SMEFT operators in the broken
phase were worked out in~\cite{Jenkins:2017jig}, and here we review
their main results relevant to this paper, limiting ourselves to the
gauge sector.

When including dimension-6 operators, the kinetic terms of the gauge
bosons receive contributions from the SMEFT operators $(H^{\dagger}H)W_{\mu\nu}^{I}W^{I\mu\nu}$
and $(H^{\dagger}H)B_{\mu\nu}B^{\mu\nu}$, and thus must be normalized.
In order to do so, the gauge fields and couplings in the unbroken
phase are redefined as follows: 
\begin{align}
W_{\mu}^{I} & =\mathcal{W}_{\mu}^{I}(1+C_{HW}v_{T}^{2}),~~B_{\mu}=\mathcal{B}_{\mu}(1+C_{HB}v_{T}^{2}),\nonumber \\
\bar{g}_{1} & =g_{1}(1+C_{HB}v_{T}^{2}),~~\bar{g}_{2}=g_{2}(1+C_{HW}v_{T}^{2}),\label{eq:Redefinitions}
\end{align}
where $C_{HW}$ and $C_{HB}$ are the Wilson coefficients of the above
operators, and $v_{T}$ is the total Higgs vacuum expectation value
(VEV) including all contributions from dim-6 operators that contribute
to the Higgs potential~\footnote{The dim-6 operators that contribute to the Higgs VEV and kinetic terms
are $\mathcal{L}_{H}^{6}=C_{H}(H^{\dagger}H)^{3}+C_{H\Box}(H^{\dagger}H)\Box(H^{\dagger}H)+C_{HD}(H^{\dagger}D^{\mu}H)^{*}(H^{\dagger}D_{\mu}H)$,
where the first one contributes to the VEV, whereas the remaining
two contribute to the kinetic term.}, which can be expressed as 
\begin{equation}
v_{T}=\Big(1+\frac{3C_{H}v^{2}}{8\lambda}\Big)v.\label{eq:total_vev}
\end{equation}
$\mathcal{W}_{\mu}^{I}$ and $\mathcal{B}_{\mu}$ can be rotated into
the physical states as follows 
\begin{equation}
\begin{pmatrix}\mathcal{Z}^{\mu}\\
\mathcal{A}^{\mu}
\end{pmatrix}=\begin{pmatrix}\bar{c}-\frac{\epsilon}{2}\bar{s}\hspace{5mm} & -\bar{s}+\frac{\epsilon}{2}\bar{c}\\
\bar{s}+\frac{\epsilon}{2}\bar{c}\hspace{5mm} & \bar{c}+\frac{\epsilon}{2}\bar{s}
\end{pmatrix}\begin{pmatrix}\mathcal{W}_{3}^{\mu}\\
\mathcal{B}^{\mu}
\end{pmatrix},\label{eq:rotation}
\end{equation}
where $\epsilon=C_{HWB}v_{T}^{2}$, with $C_{HWB}$ being the coefficient
of the SMEFT operator $(H^{\dagger}\sigma^{I}H)W_{\mu\nu}^{I}B^{\mu\nu}$,
and 
\begin{align}
\bar{c} & =\cos\bar{\theta}=\frac{\bar{g}_{2}}{\sqrt{\bar{g}_{1}^{2}+\bar{g}_{2}^{2}}}\Bigg[1-\frac{\epsilon}{2}\frac{\bar{g}_{1}}{\bar{g}_{2}}\Bigg(\frac{\bar{g}_{2}^{2}-\bar{g}_{1}^{2}}{\bar{g}_{2}^{2}+\bar{g}_{1}^{2}}\Bigg)\Bigg],\nonumber \\
\bar{s} & =\sin\bar{\theta}=\frac{\bar{g}_{1}}{\sqrt{\bar{g}_{1}^{2}+\bar{g}_{2}^{2}}}\Bigg[1+\frac{\epsilon}{2}\frac{\bar{g}_{2}}{\bar{g}_{1}}\Bigg(\frac{\bar{g}_{2}^{2}-\bar{g}_{1}^{2}}{\bar{g}_{2}^{2}+\bar{g}_{1}^{2}}\Bigg)\Bigg].\label{eq:Mod_c,s}
\end{align}
Furthermore, the mass and effective couplings of the physical $\mathcal{Z}_{\mu}$
get modified: 
\begin{align}
M_{\mathcal{Z}}^{2} & =\frac{1}{4}(\bar{g}_{1}^{2}+\bar{g}_{2}^{2})v_{T}^{2}\Big(1+\frac{1}{2}C_{HD}v_{T}^{2}\Big)+\frac{\epsilon}{2}\bar{g}_{1}\bar{g}_{2}v_{T}^{2},\nonumber \\
\bar{g}_{Z} & =\frac{\bar{e}}{\sin{\bar{\theta}}\cos{\bar{\theta}}}\Bigg[1+\Bigg(\frac{\bar{g}_{1}^{2}+\bar{g}_{2}^{2}}{2\bar{g}_{1}\bar{g}_{2}}\Bigg)v_{T}^{2}C_{HWB}\Bigg],\nonumber \\
\bar{e} & =\bar{g}_{2}\sin{\bar{\theta}}-\frac{1}{2}\cos{\bar{\theta}}\bar{g}_{2}v_{T}^{2}C_{HWB}.\label{eq:Z_mass}
\end{align}

Now, we have all the ingredients needed to extract the couplings of
the physical $\mathcal{Z}$ to the rest of the SM particles. For fermions,
$\mathcal{Z}ff$ can be expressed as 
\begin{equation}
\mathcal{L}_{\mathcal{Z}}^{(6)}=-\bar{g}_{Z}\mathcal{Z}_{\mu}J_{\mathcal{Z}}^{\mu},\label{eq:BSm_Z_couplings}
\end{equation}
where neglecting neutrinos, the neutral current $J_{\mathcal{Z}}^{\mu}$
is given by 
\begin{equation}
\begin{aligned}J_{\mathcal{Z}}^{\mu} & =[Z_{e_{L}}]_{ij}\bar{e}_{L_{i}}\gamma^{\mu}e_{L_{j}}+[Z_{e_{R}}]_{ij}\bar{e}_{R_{i}}\gamma^{\mu}e_{R_{j}}+[Z_{u_{L}}]_{ij}\bar{u}_{L_{i}}\gamma^{\mu}u_{L_{j}}\\
 & +[Z_{u_{R}}]_{ij}\bar{u}_{R_{i}}\gamma^{\mu}u_{R_{j}}+[Z_{d_{L}}]_{ij}\bar{d}_{L_{i}}\gamma^{\mu}d_{L_{j}}+[Z_{d_{R}}]_{ij}\bar{d}_{R_{i}}\gamma^{\mu}d_{R_{j}},
\end{aligned}
\label{eq:BSM_neutral_current}
\end{equation}
where $i,j$ are generation indices, and the coefficients are given
by 
\begin{equation}
\begin{aligned}[Z]_{ij} & =\Bigg[\Big(-\frac{1}{2}+\bar{s}^{2}\Big)\delta_{ij}-\frac{1}{2}v_{T}^{2}C_{\substack{Hl\\
ij
}
}^{(1)}-\frac{1}{2}v_{T}^{2}C_{\substack{Hl\\
ij
}
}^{(3)}\Bigg],\\{}
[Z_{e_{R}}]_{ij} & =\Bigg[\bar{s}^{2}\delta_{ij}-\frac{1}{2}v_{T}^{2}C_{\substack{He\\
ij
}
}\Bigg],\\{}
[Z_{u_{L}}]_{ij} & =\Bigg[\Big(\frac{1}{2}-\frac{2}{3}\bar{s}^{2}\Big)\delta_{ij}-\frac{1}{2}v_{T}^{2}C_{\substack{Hq\\
ij
}
}^{(1)}+\frac{1}{2}v_{T}^{2}C_{\substack{Hq\\
ij
}
}^{(3)}\Bigg],\\{}
[Z_{u_{R}}]_{ij} & =\Bigg[-\frac{2}{3}\bar{s}^{2}\delta_{ij}-\frac{1}{2}v_{T}^{2}C_{\substack{Hu\\
ij
}
}\Bigg],\\{}
[Z_{d_{L}}]_{ij} & =\Bigg[\Big(-\frac{1}{2}+\frac{1}{3}\bar{s}^{2}\Big)\delta_{ij}-\frac{1}{2}v_{T}^{2}C_{\substack{Hq\\
ij
}
}^{(1)}-\frac{1}{2}v_{T}^{2}C_{\substack{Hq\\
ij
}
}^{(3)}\Bigg],\\{}
[Z_{d_{R}}]_{ij} & =\Bigg[\frac{1}{3}\bar{s}^{2}\delta_{ij}-\frac{1}{2}v_{T}^{2}C_{\substack{Hd\\
ij
}
}\Bigg],
\end{aligned}
\label{eq:BSM_coefficients}
\end{equation}
and as usual, the FV coefficients are given by $i\neq j$, whereas
the flavor-conserving ones, $i=j$ contribute to the SM couplings.
Thus, it is possible to tune the diagonal contributions to zero while
allowing the off-diagonal ones to be free parameters via an appropriate
choice of the parameters. For example, if we set $C_{H}=C_{HB}=C_{HW}=C_{HWB}=0$,
then this implies that $v_{T}=v$, $\bar{g}_{1,2}=g_{1,2}$, $\bar{g}_{Z}=g_{Z}$,
$\bar{\theta}=\theta_{W}$, $\mathcal{Z}^{\mu}=Z^{\mu}$, $\mathcal{A}^{\mu}=A^{\mu}$
and $M_{\mathcal{Z}}=m_{Z}$, and setting $C_{\substack{xx\\
ii
}
}=0$ while allowing $C_{\substack{xx\\
ij
}
}\neq0$ for $i\ne j$ will introduce no contributions to the diagonal terms,
while inducing non-vanishing contributions to the off-diagonal couplings.
As a concrete example, consider the $Z\bar{e}e$ interaction 
\begin{equation}
\mathcal{L}_{ee}=-g_{Z}Z_{\alpha}\Big([Z_{e_{L}}]_{ij}\bar{e}_{L_{i}}\gamma^{\alpha}e_{L_{j}}+[Z_{e_{R}}]_{ij}\bar{e}_{R_{i}}\gamma^{\alpha}e_{R_{j}}\Big),\label{eq:Zee}
\end{equation}
with the coefficients given by the first and second lines of Eq.~(\ref{eq:BSM_coefficients}).
If we set $C_{\substack{Hl\\
11
}
}^{(1)}=C_{\substack{Hl\\
11
}
}^{(3)}=C_{\substack{He\\
11
}
}=0$, then $[Z_{e_{L}}]_{11}$ and $[Z_{e_{R}}]_{11}$ will reduce to
their SM values and the diagonal part of Eq.~(\ref{eq:Zee}) reduces
to the SM Lagrangian. On the other hand, allowing $C_{\substack{Hl\\
1i
}
}^{(1)}$, $C_{\substack{Hl\\
1i
}
}^{(3)}$, and $C_{\substack{He\\
1i
}
}$ (for $i=2,3$) to have non-vanishing values will give rise to the
off-diagonal interaction term 
\begin{equation}
\mathcal{L}_{e\mu,e\tau}=-g_{Z}Z_{\alpha}\Big([Z_{e_{L}}]_{12}\bar{e}_{L}\gamma^{\alpha}\mu_{L}+[Z_{e_{R}}]_{12}\bar{e}_{R}\gamma^{\alpha}\mu_{R}+H.c.\Big)+(\mu \to \tau),\label{eq:FV_lepton_SMEFT}
\end{equation}
which can immediately be matched to Eq.~(\ref{eq:FV_Z_coupling}).
Another possibility is to simply absorb all diagonal dim-6 contributions
in the definition of the SM couplings, however, this requires the
assumption that such contributions are all flavor-universal or very
close to flavor-universal as is the case in the SM. Thus to summarize,
it is possible to suppress the contributions of higher-dimensional
operators to the diagonal terms of $g_{L,R}^{ij}$ while allowing
the off-diagonal ones to be sizable, albeit at the expense of some
fine-tuning that will always be unavoidable.~\footnote{We also point out that RGE flows will ruin any tuning in the UV, however,
this is irrelevant in the IR region where we are interested.}

\begin{table}[ht]
\centering \vspace{1mm}
 \tabcolsep7pt
\begin{tabular}{lccc}
\hline 
\textbf{Channel} & \textbf{Couplings} & \textbf{Bounds} & \textbf{Projections} \tabularnewline
\hline 
\hline 
$Z\to\tau\mu$ & $\sqrt{|g_{L}^{\tau\mu}|^{2}+|g_{R}^{\tau\mu}|^{2}}$ & $<2.12\times10^{-3}$ & $<2.91\times10^{-5}$ \tabularnewline
$\text{EWPO - \ensuremath{U} parameter}$ & $\sqrt{|g_{L}^{\tau\mu}|^{2}+|g_{R}^{\tau\mu}|^{2}}$ & $<0.2$ & $-$ \tabularnewline
$Z\to\mu^{+}\mu^{-}$ & $\sqrt{g_{L}^{2}|g_{L}^{\tau\mu}|^{2}+g_{R}^{2}|g_{R}^{\tau\mu}|^{2}}$ & $<0.11$ & $-$ \tabularnewline
$Z\to\tau^{+}\tau^{-}$ & $\sqrt{g_{L}^{2}|g_{L}^{\tau\mu}|^{2}+g_{R}^{2}|g_{R}^{\tau\mu}|^{2}}$ & $<0.12$ & $-$ \tabularnewline
$\tau\to\mu\gamma$ & $\sqrt{g_{L}^{2}|g_{L}^{\tau\mu}|^{2}+g_{R}^{2}|g_{R}^{\tau\mu}|^{2}}$ & $<2.85\times10^{-7}$ & $<4.4\times10^{-8}$ \tabularnewline
$\tau\to3\mu$ & $\sqrt{g_{L}^{2}|g_{L}^{\tau\mu}|^{2}+g_{R}^{2}|g_{R}^{\tau\mu}|^{2}}$ & $<8.64\times10^{-7}$ & $<1.33\times10^{-7}$ \tabularnewline
$\tau\to\mu^{-}e^{+}e^{-}$ & $\sqrt{g_{L}^{2}|g_{L}^{\tau\mu}|^{2}+g_{R}^{2}|g_{R}^{\tau\mu}|^{2}}$ & $<5.66\times10^{-7}$ & - \tabularnewline
$\tau\to\mu^{+}e^{-}e^{-}$ & $\sqrt{g_{L}^{2}|g_{L}^{\tau\mu}|^{2}+g_{R}^{2}|g_{R}^{\tau\mu}|^{2}}$ & $<7.3\times10^{-7}$ & - \tabularnewline
$\Upsilon(1S)\to\tau\mu$ & $\sqrt{|g_{L}^{\tau\mu}|^{2}+|g_{R}^{\tau\mu}|^{2}}$ & $<0.11$ & - \tabularnewline
$\Upsilon(2S)\to\tau\mu$ & $\sqrt{|g_{L}^{\tau\mu}|^{2}+|g_{R}^{\tau\mu}|^{2}}$ & $<7.81\times10^{-2}$ & - \tabularnewline
$\Upsilon(3S)\to\tau\mu$ & $\sqrt{|g_{L}^{\tau\mu}|^{2}+|g_{R}^{\tau\mu}|^{2}}$ & $<5.11\times10^{-2}$ & - \tabularnewline
$J/\psi\to\tau\mu$ & $\sqrt{|g_{L}^{\tau\mu}|^{2}+|g_{R}^{\tau\mu}|^{2}}$ & $<0.95$ & - \tabularnewline
$B^{0}\to\tau\mu$ & $\sqrt{|g_{L}^{\tau\mu}|^{2}+|g_{R}^{\tau\mu}|^{2}}$ & $<3.42\times10^{-4}$ & - \tabularnewline
$B_{s}\to\tau\mu$ & $\sqrt{|g_{L}^{\tau\mu}|^{2}+|g_{R}^{\tau\mu}|^{2}}$ & $<7.24\times10^{-3}$ & - \tabularnewline
$\tau\to\mu\rho^{0}$ & $\sqrt{|g_{L}^{\tau\mu}|^{2}+|g_{R}^{\tau\mu}|^{2}}$ & $<5.5\times10^{-5}$ & - \tabularnewline
$\tau\to\mu\omega$ & $\sqrt{|g_{L}^{\tau\mu}|^{2}+|g_{R}^{\tau\mu}|^{2}}$ & $<8.34\times10^{-5}$ & - \tabularnewline
$\tau\to\mu\phi$ & $\sqrt{|g_{L}^{\tau\mu}|^{2}+|g_{R}^{\tau\mu}|^{2}}$ & $<6.12\times10^{-5}$ & - \tabularnewline
$\tau\to\mu\pi^{0}$ & $\sqrt{|g_{L}^{\tau\mu}|^{2}+|g_{R}^{\tau\mu}|^{2}}$ & $<2.15\times10^{-4}$ & - \tabularnewline
$\tau\to\mu K_{S}^{0}$ & $\sqrt{|g_{L}^{\tau\mu}|^{2}+|g_{R}^{\tau\mu}|^{2}}$ & $<8.92\times10^{-5}$ & - \tabularnewline
$\tau\to\mu\eta$ & $\sqrt{|g_{L}^{\tau\mu}|^{2}+|g_{R}^{\tau\mu}|^{2}}$ & $<2.19\times10^{-4}$ & $2.72\times10^{-5}$ \tabularnewline
$\tau\to\mu\eta'$ & $\sqrt{|g_{L}^{\tau\mu}|^{2}+|g_{R}^{\tau\mu}|^{2}}$ & $<4.8\times10^{-4}$ & -\tabularnewline
\hline 
Michel Parameters & $\sqrt{|g_{L}^{\tau\mu}|^{2}+|g_{R}^{\tau\mu}|^{2}+|g_{L}^{\tau e}|^{2}+|g_{R}^{\tau e}|^{2}}$ & $<6.52\times10^{-2}$ & - \tabularnewline
\hline 
\end{tabular}\caption{ {\small{}{}{}The $90\%$ CL bounds and projections on the leptonic
next-to-minimal FV $Z$ couplings to $\tau$ and $\mu$.}}
\label{table1} 
\end{table}

\section{Bounds on The FV $Z$ Interactions in the Leptonic Sector}

~\label{sec:3}

In this section, we will discuss the relevant constraints on the FV
$Z$ couplings to leptons. We have investigated FV in the quark sector
in~\cite{Abu-Ajamieh:2025jov}. Explicitly, we write the leptonic
part of Eq.~(\ref{eq:FV_Z_coupling}) as follows 
\begin{equation}
\mathcal{L}_{l}=\mathcal{L}_{l}^{\text{SM}}-Z_{\alpha}\bar{\tau}\gamma^{\alpha}(g_{L}^{\tau\mu}P_{L}+g_{R}^{\tau\mu}P_{R})\mu-Z_{\alpha}\bar{\tau}\gamma^{\alpha}(g_{L}^{\tau e}P_{L}+g_{R}^{\tau e}P_{R})e-Z_{\alpha}\bar{\mu}\gamma^{\alpha}(g_{L}^{\mu e}P_{L}+g_{R}^{\mu e}P_{R})e+\text{h.c.}\label{eq:FV_lepton_Lag}
\end{equation}

This is the main formula we use throughout this paper to extract bounds
on the FV $Z$ couplings to leptons. All the bounds reported in this
paper are given $@90\%$ CL. Tables~\ref{table1},~\ref{table2}
and~\ref{table3} summarize the results, in addition to the plots
in Fig.~\ref{fig10}.

\begin{table}[ht]
\centering \vspace{1mm}
 \tabcolsep7pt%
\begin{tabular}{lccc}
\hline 
\textbf{Channel} & \textbf{Couplings} & \textbf{Bounds} & \textbf{Projections} \tabularnewline
\hline 
\hline 
$Z\to\tau e$ & $\sqrt{|g_{L}^{\tau e}|^{2}+|g_{R}^{\tau e}|^{2}}$ & $<2.06\times10^{-3}$ & $<2.91\times10^{-5}$ \tabularnewline
$\text{EWPO - \ensuremath{U} parameter}$ & $\sqrt{|g_{L}^{\tau e}|^{2}+|g_{R}^{\tau e}|^{2}}$ & $<0.2$ & $-$ \tabularnewline
$\text{LEP}(e^{+}e^{-}\to\tau^{+}\tau^{-})$ & $\sqrt{|g_{L}^{\tau e}|^{2}+0.91|g_{R}^{\tau e}|^{2}}$ & < $6.28\times10^{-2}$ & $-$ \tabularnewline
$Z\to e^{+}e^{-}$ & $\sqrt{g_{L}^{2}|g_{L}^{\tau e}|^{2}+g_{R}^{2}|g_{R}^{\tau e}|^{2}}$ & $<8.43\times10^{-2}$ & $-$ \tabularnewline
$Z\to\tau^{+}\tau^{-}$ & $\sqrt{g_{L}^{2}|g_{L}^{\tau e}|^{2}+g_{R}^{2}|g_{R}^{\tau e}|^{2}}$ & $<0.12$ & $-$ \tabularnewline
$\tau\to e\gamma$ & $\sqrt{g_{L}^{2}|g_{L}^{\tau e}|^{2}+g_{R}^{2}|g_{R}^{\tau e}|^{2}}$ & $<2.53\times10^{-7}$ & $<9.84\times10^{-8}$ \tabularnewline
$\tau\to3e$ & $\sqrt{g_{L}^{2}|g_{L}^{\tau e}|^{2}+g_{R}^{2}|g_{R}^{\tau e}|^{2}}$ & $<4.72\times10^{-9}$ & $<6.43\times10^{-10}$ \tabularnewline
$\tau\to\mu^{-}\mu^{+}e^{-}$ & $\sqrt{g_{L}^{2}|g_{L}^{\tau e}|^{2}+g_{R}^{2}|g_{R}^{\tau e}|^{2}}$ & $<3.34\times10^{-9}$ & - \tabularnewline
$\tau\to\mu^{-}\mu^{-}e^{+}$ & $\sqrt{g_{L}^{2}|g_{L}^{\tau e}|^{2}+g_{R}^{2}|g_{R}^{\tau e}|^{2}}$ & $<3.75\times10^{-9}$ & - \tabularnewline
%$\tau\to e+\text{inv.}$ & $\sqrt{|g_{L}^{\tau e}|^{2}+|g_{R}^{\tau e}|^{2}}$ & $<2.76\times 10^{-4}$ & - \tabularnewline
$\Upsilon(2S)\to\tau e$ & $\sqrt{|g_{L}^{\tau e}|^{2}+|g_{R}^{\tau e}|^{2}}$ & $<7.69\times10^{-2}$ & - \tabularnewline
$\Upsilon(3S)\to\tau e$ & $\sqrt{|g_{L}^{\tau e}|^{2}+|g_{R}^{\tau e}|^{2}}$ & $<5.95\times10^{-2}$ & - \tabularnewline
$J/\psi\to\tau e$ & $\sqrt{|g_{L}^{\tau e}|^{2}+|g_{R}^{\tau e}|^{2}}$ & $<1.93$ & - \tabularnewline
$B^{0}\to\tau e$ & $\sqrt{|g_{L}^{\tau e}|^{2}+|g_{R}^{\tau e}|^{2}}$ & $<3.66\times10^{-4}$ & - \tabularnewline
$B_{s}\to\tau e$ & $\sqrt{|g_{L}^{\tau e}|^{2}+|g_{R}^{\tau e}|^{2}}$ & $<4.18\times10^{-2}$ & - \tabularnewline
$\tau\to e\rho^{0}$ & $\sqrt{|g_{L}^{\tau e}|^{2}+|g_{R}^{\tau e}|^{2}}$ & $<6.26\times10^{-5}$ & - \tabularnewline
$\tau\to e\omega$ & $\sqrt{|g_{L}^{\tau e}|^{2}+|g_{R}^{\tau e}|^{2}}$ & $<6.55\times10^{-5}$ & - \tabularnewline
$\tau\to e\phi$ & $\sqrt{|g_{L}^{\tau e}|^{2}+|g_{R}^{\tau e}|^{2}}$ & $<5.71\times10^{-5}$ & - \tabularnewline
$\tau\to e\pi^{0}$ & $\sqrt{|g_{L}^{\tau e}|^{2}+|g_{R}^{\tau e}|^{2}}$ & $<1.83\times10^{-4}$ & - \tabularnewline
$\tau\to eK_{S}^{0}$ & $\sqrt{|g_{L}^{\tau e}|^{2}+|g_{R}^{\tau e}|^{2}}$ & $<9.48\times10^{-5}$ & - \tabularnewline
$\tau\to e\eta$ & $\sqrt{|g_{L}^{\tau e}|^{2}+|g_{R}^{\tau e}|^{2}}$ & $<2.61\times10^{-4}$ & - \tabularnewline
$\tau\to e\eta'$ & $\sqrt{|g_{L}^{\tau e}|^{2}+|g_{R}^{\tau e}|^{2}}$ & $<5.32\times10^{-4}$ & - \tabularnewline
\hline 
\end{tabular}\caption{{\small{}{}{}The $90\%$ CL bounds and projections on the leptonic
next-to-minimal FV $Z$ couplings to $\tau$ and $e$.}}
\label{table2} 
\end{table}

\begin{table}[ht]
\centering \vspace{1mm}
 \tabcolsep7pt%
\begin{tabular}{lccc}
\hline 
\textbf{Channel} & \textbf{Couplings} & \textbf{Bounds} & \textbf{Projections} \tabularnewline
\hline 
\hline 
$Z\to\mu e$ & $\sqrt{|g_{L}^{\mu e}|^{2}+|g_{R}^{\mu e}|^{2}}$ & $<4.02\times10^{-4}$ & $<9.21\times10^{-6}$ \tabularnewline
$\text{EWPO - \ensuremath{U} parameter}$ & $\sqrt{|g_{L}^{\mu e}|^{2}+|g_{R}^{\mu e}|^{2}}$ & $<0.2$ & $-$ \tabularnewline
$\text{LEP}(e^{+}e^{-}\to\mu^{+}\mu^{-})$ & $\sqrt{|g_{L}^{\mu e}|^{2}+0.91|g_{R}^{\mu e}|^{2}}$ & < $5.62\times10^{-2}$ & $-$ \tabularnewline
$Z\to e^{+}e^{-}$ & $\sqrt{g_{L}^{2}|g_{L}^{\mu e}|^{2}+g_{R}^{2}|g_{R}^{\mu e}|^{2}}$ & $<8.43\times10^{-2}$ & $-$ \tabularnewline
$Z\to\mu^{+}\mu^{-}$ & $\sqrt{g_{L}^{2}|g_{L}^{\mu e}|^{2}+g_{R}^{2}|g_{R}^{\mu e}|^{2}}$ & $<0.11$ & $-$ \tabularnewline
$\mu\to e\gamma$ & $\sqrt{g_{L}^{2}|g_{L}^{\mu e}|^{2}+g_{R}^{2}|g_{R}^{\mu e}|^{2}}$ & $<1.34\times10^{-12}$ & $<4.63\times10^{-13}$ \tabularnewline
$\mu\to3e$ & $\sqrt{g_{L}^{2}|g_{L}^{\mu e}|^{2}+g_{R}^{2}|g_{R}^{\mu e}|^{2}}$ & $<7.17\times10^{-13}$ & $<7.17\times10^{-15}$ \tabularnewline
$\mu\to e\hspace{2mm}\text{conversion}$ & $\sqrt{|g_{L}^{\mu e}|^{2}+0.74|g_{R}^{\mu e}|^{2}}$ & $<5.84\times10^{-11}$ & $<9.49\times10^{-13}$ \tabularnewline
%$\mu\to e+\text{inv.}$ & $\sqrt{|g_{L}^{\mu e}|^{2}+|g_{R}^{\mu e}|^{2}}$ & $<8.51\times 10^{-8}$ & $<1.12\times 10^{-9}$ \tabularnewline
$\mu\hspace{2mm}\text{lifetime}$ & $\sqrt{|g_{L}^{\mu e}|^{2}+|g_{R}^{\mu e}|^{2}}$ & $<9.45\times10^{-4}$ & - \tabularnewline
$M-\overline{M}\hspace{2mm}\text{oscillation}$ & $(|g_{L}^{\mu e}|^{2}+|g_{R}^{\mu e}|^{2})^{2}+6.5|g_{L}^{\mu e}|^{2}|g_{R}^{\mu e}|^{2}$ & $<2.72\times10^{-6}$ & - \tabularnewline
 & $-2.32(|g_{L}^{\mu e}|^{2}+|g_{R}^{\mu e}|^{2})\text{Re}(g_{L}^{\mu e}g_{R}^{\mu e*})$ & & \tabularnewline
$J/\psi\to\mu e$ & $\sqrt{|g_{L}^{\mu e}|^{2}+|g_{R}^{\mu e}|^{2}}$ & $<0.19$ & - \tabularnewline
$\phi\to\mu e$ & $\sqrt{|g_{L}^{\mu e}|^{2}+|g_{R}^{\mu e}|^{2}}$ & $<42.93$ & - \tabularnewline
$\eta\to\mu e$ & $\sqrt{|g_{L}^{\mu e}|^{2}+|g_{R}^{\mu e}|^{2}}$ & $<22.74$ & - \tabularnewline
$\eta'\to\mu e$ & $\sqrt{|g_{L}^{\mu e}|^{2}+|g_{R}^{\mu e}|^{2}}$ & $<2.16\times10^{3}$ & - \tabularnewline
$\pi^{0}\to\mu e$ & $\sqrt{|g_{L}^{\mu e}|^{2}+|g_{R}^{\mu e}|^{2}}$ & $<5.7\times10^{-2}$ & - \tabularnewline
$B^{0}\to\mu e$ & $\sqrt{|g_{L}^{\mu e}|^{2}+|g_{R}^{\mu e}|^{2}}$ & $<4.3\times10^{-5}$ & - \tabularnewline
$B_{s}\to\mu e$ & $\sqrt{|g_{L}^{\mu e}|^{2}+|g_{R}^{\mu e}|^{2}}$ & $<8.21\times10^{-5}$ & - \tabularnewline
$D^{0}\to\mu e$ & $\sqrt{|g_{L}^{\mu e}|^{2}+|g_{R}^{\mu e}|^{2}}$ & $<4.51\times10^{-4}$ & - \tabularnewline
$K_{L}^{0}\to\mu e$ & $\sqrt{|g_{L}^{\mu e}|^{2}+|g_{R}^{\mu e}|^{2}}$ & $<6.57\times10^{-8}$ & - \tabularnewline
Michel Parameters & $\sqrt{|g_{L}^{\mu e}|^{2}+|g_{R}^{\mu e}|^{2}}$ & $<7.99\times10^{-3}$ & - \tabularnewline
\hline 
\end{tabular}\caption{ {\small{}{}{}The $90\%$ CL bounds and projections on the leptonic
FV $Z$ couplings to $\mu$ and $e$.}}
\label{table3} 
\end{table}

\subsection{Constraints from $Z\to\ell_{i}^{\pm}\ell_{j}^{\mp}$ Direct Searches}

The first and most straightforward bound arises from the direct searches
for FV $Z$ decays, which have been updated recently by the LHC. This
process can take place at tree-level and a simple calculation yields
\begin{equation}
\Gamma(Z\to\ell_{i}^{\pm}\ell_{j}^{\mp})\simeq\frac{m_{Z}}{12\pi}\Big(|g_{L}^{ij}|^{2}+|g_{R}^{ij}|^{2}\Big),\label{eq:Z_decay_ij}
\end{equation}
where $\ell_{i,j}=\{e,\mu,\tau\}$ and we have dropped terms of $\mathcal{O}(\frac{m_{\ell_{i,j}}^{2}}{M_{Z}^{2}})$.
Notice here that there is a factor of 2 stemming from the hermitian
conjugate part of the Lagrangian, since the experimental bound is
given for $Z\to\ell^{\pm}\ell^{\mp}$. For the decay $Z\to\mu^{\pm}e^{\mp}$,
the latest bounds can be found in~\cite{CMS:2025wqy} (see also~\cite{ATLAS:2022uhq})
and read $\text{Br}(Z\to\mu^{\pm}e^{\mp})<1.9\times10^{-7}$ $@95\%$
CL. This can be translated into the $90\%$ CL bound~\footnote{Whenever bounds are reported at a CL different from $90\%$, they
are shifted to $90\%$ to maintain consistency.} $\sqrt{|g_{L}^{\mu e}|^{2}+|g_{R}^{\mu e}|^{2}}<4.02\times10^{-4}$.
On the other hand, the bounds on $Z\to\tau^{\pm}\mu^{\mp},\tau^{\pm}e^{\mp}$
can be found in~\cite{CMS:2025wqy,ATLAS:2021bdj} and yield (also
$@95\%$ CL) $\text{Br}(Z\to\tau^{\pm}\mu^{\mp})<5.3\times10^{-6}$
and $\text{Br}(Z\to\tau^{\pm}e^{\mp})<5\times10^{-6}$. These translate
to $\sqrt{|g_{L}^{\tau\mu}|^{2}+|g_{R}^{\tau\mu}|^{2}}<2.12\times10^{-3}$
and $\sqrt{|g_{L}^{\tau e}|^{2}+|g_{R}^{\tau e}|^{2}}<2.06\times10^{-3}$,
respectively.

The proposed FCC-ee collider measurements can improve the bounds on
the FV $Z$ decay to $\ell_{i}^{\pm}\ell_{j}^{\mp}$. The projections
$@95\%$ CL are estimated to be $\text{Br}(Z\to\mu^{\pm}e^{\mp})\simeq10^{-8}-10^{-10}$
and $\text{Br}(Z\to\tau^{\pm}e^{\mp}/\tau^{\pm}\mu^{\mp})\simeq10^{-9}$~\cite{Calibbi:2021pyh}.
These projections translate into the projected limits $\sqrt{|g_{L}^{\mu e}|^{2}+|g_{R}^{\mu e}|^{2}}<9.21\times10^{-6}$,
$\sqrt{|g_{L}^{\tau\mu}|^{2}+|g_{R}^{\tau\mu}|^{2}}<2.91\times10^{-5}$
and $\sqrt{|g_{L}^{\tau e}|^{2}+|g_{R}^{\tau e}|^{2}}<2.91\times10^{-5}$,
respectively, where we have used the strongest projection to set the
limits on the FV $Z$ couplings to $\mu e$.

\subsection{LEP Constraints}

\begin{figure}[b!]
\includegraphics[width=0.65\textwidth]{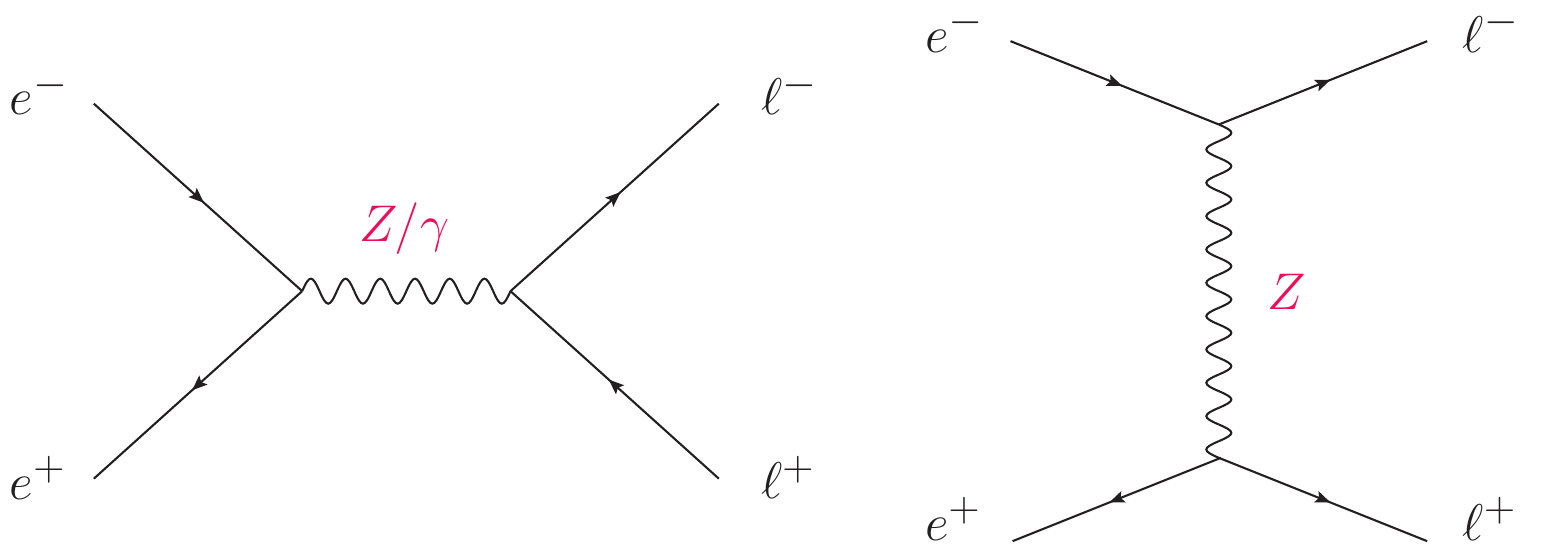} \caption{LEP searches involving $e^{+}e^{-}\to\mu^{+}\mu^{-},\tau^{+}\tau^{-}$.
Only the t-channel could have FV couplings.}
\label{fig1} 
\end{figure}

LEP searches involving $e^{+}e^{-}\to\mu^{+}\mu^{-},\tau^{+}\tau^{-}$
can be used to set limits on the FV $Z$ couplings. These processes
proceed via the s-channel and t-channel as shown in Fig.~\ref{fig1},
with only the t-channel ones being FV.

To find the cross-section, we express the total matrix element as
follows 
\begin{equation}
\mathcal{M}=\mathcal{M}_{\text{SM}}+\mathcal{M}_{\text{FV}},\label{eq:LEP_mat}
\end{equation}
where $\mathcal{M}_{\text{SM}}$ ($\mathcal{M}_{\text{FV}}$) is given
by the s- (t-) channel. Working to the LO in $g_{L,R}^{el}$, it is
easy to see that the leading FV contribution arises from the cross
terms between the SM and the FV contributions $2\text{Re}(\mathcal{M}_{\text{SM}}\mathcal{M}_{\text{FV}}^{*})$
and is given by 
\begin{equation}
\begin{aligned}\delta\sigma_{\text{FV}}(e^{+}e^{-}\to\ell^{-}\ell^{+}) & \simeq\frac{1}{16\pi s}\Bigg[2\left(1+\frac{m_{Z}^{2}}{s}\right)^{2}\log\left(1+\frac{s}{m_{Z}^{2}}\right)-\left(3+\frac{2m_{Z}^{2}}{s}\right)\Bigg]\times\\
 & \left(\left(g_{L}^{2}\frac{s}{s-m_{Z}^{2}}+4\pi\alpha\right)|g_{L}^{e\ell}|^{2}+\left(g_{R}^{2}\frac{s}{s-m_{Z}^{2}}+4\pi\alpha\right)|g_{R}^{e\ell}|^{2}\right).
\end{aligned}
\label{eq:LEP_Xsection}
\end{equation}

This can be used to extract bounds on the FV $Z$ couplings by demanding
that the FV cross-sections in Eq.~(\ref{eq:LEP_Xsection}) be within
the measured experimental uncertainties. The measured total cross-section
of these processes can be found in~\cite{ALEPH:2006bhb}, where we
find that the $1\sigma$ uncertainties on the cross-sections of $e^{+}e^{-}\to\mu^{+}\mu^{-},\tau^{+}\tau^{-}$
are $0.088$ pb and $0.11$ pb, respectively at a (Center-of-Momentum)
COM energy of $\sqrt{s}=207$ GeV. These measurements translate to
the bounds $\sqrt{|g_{L}^{\mu e}|^{2}+0.91|g_{R}^{\mu e}|^{2}}<5.62\times10^{-2}$
and $\sqrt{|g_{L}^{\tau e}|^{2}+0.91|g_{R}^{\tau e}|^{2}}<6.28\times10^{-2}$.
We point out that future $e^{+}e^{-}$ colliders, such as International
Linear Collider (ILC)~\cite{LinearColliderVision:2025hlt} and the
Future Circular Electron-Position Collider (FCC-ee)~\cite{dEnterria:2016sca}
could potentially improve these limits.~\footnote{As the LEP bound is rather weak compared to the other bounds, using
other measured observables from LEP will not improve the bound by
much to make it comparable to other stronger bounds, such as from
$\ell_{i}\to 3\ell_{j}$ or $\ell_{i}\to \ell_{j}\gamma$.}

\subsection{Constraints from Loop Correction to the $Z\to\ell_{i}^{+}\ell_{i}^{-}$
Decay}

\begin{figure}[t!]
\includegraphics[width=0.8\textwidth]{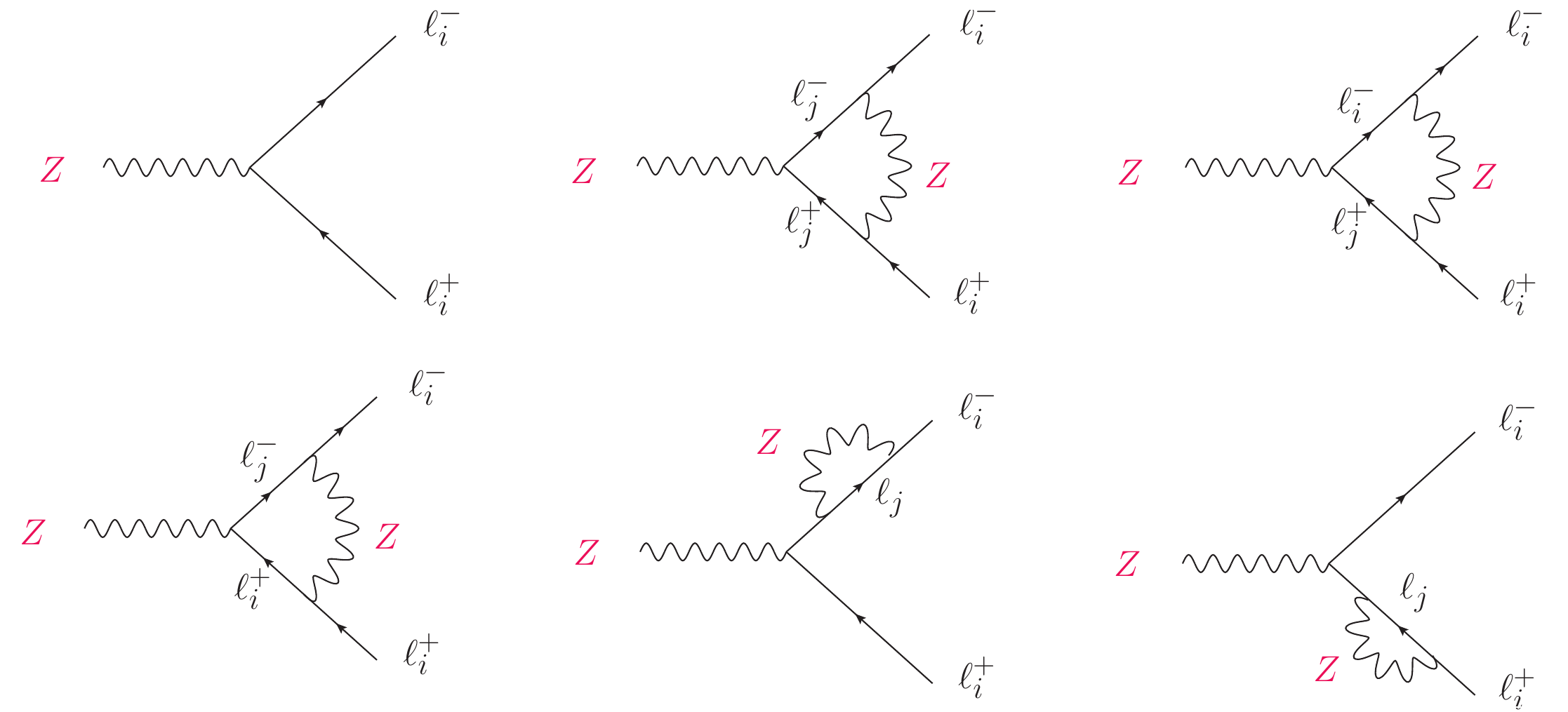} \caption{Tree level and FV corrections to $Z\to\ell_{i}^{+}\ell_{i}^{-}$ at
one loop at LO in the FV couplings.}
\label{fig2} 
\end{figure}

Although the $Z$ decay to a pair of identical leptons does not have
FV couplings at tree level, it nonetheless can receive contributions
from FV couplings at 1-loop as shown in Fig. ~\ref{fig2}. Thus,
it is possible to extract bounds on the FV $Z$ couplings, especially
since the branching ratios of the leptonic $Z$ decays are well-measured.
In our calculation, we drop terms of $O(\frac{m_{i,j}^{2}}{m_{Z}^{2}})$.
There are two types of FV corrections at 1-loop: Vertex corrections
and leg corrections. As it turns out, the leading terms in the leg
corrections are of $O(\frac{m_{i,j}^{2}}{m_{Z}^{2}})$, thus we neglect
them and only consider the vertex corrections. Using dimensional regularization
in the $\overline{\text{MS}}$ scheme, the leading FV vertex corrections
read 
\begin{equation}
\Gamma(Z\to\ell_{i}^{+}\ell_{i}^{-})\simeq\Gamma_{0}+\frac{(16\pi^{2}-171)}{1536\pi^{3}}(g_{L}^{2}|g_{L}^{ij}|^{2}+g_{R}^{2}|g_{R}^{ij}|^{2})m_{Z},\label{eq:Ztoll}
\end{equation}
where $\Gamma_{0}=\frac{1}{24\pi}(g_{L}^{2}+g_{R}^{2})m_{Z}$ is the
tree-level decay width. The latest bounds on the leptonic $Z$ decays
are given by~\cite{ParticleDataGroup:2024cfk} 
\begin{align}
\text{Br}(Z\to e^{+}e^{-}) & =(3363.2\pm4.2)\times10^{-3}\%,\label{eq:ZtollExp_a}\\
\text{Br}(Z\to\mu^{+}\mu^{-}) & =(3366.2\pm6.6)\times10^{-3}\%,\label{eq:ZtollExp_b}\\
\text{Br}(Z\to\tau^{+}\tau^{-}) & =(3369.6\pm8.3)\times10^{-3}\%.\label{eq:ZtollExp_c}
\end{align}
Thus, to obtain the $90\%$ CL bound, we require 
\begin{equation}
\frac{\delta\Gamma(Z\to\ell_{i}^{+}\ell_{i}^{-})}{\Gamma_{Z}}=\frac{\Gamma(Z\to\ell_{i}^{+}\ell_{i}^{-})-\Gamma_{0}}{\Gamma_{Z}}\lesssim1.7\times\delta\text{Br}^{\text{Exp}}(Z\to\ell_{i}^{+}\ell_{i}^{-}),\label{eq:ZtollBound1}
\end{equation}
where $\Gamma_{Z}$ is the experimentally-measured $Z$ decay width,
and we have introduced a factor of 1.7 to shift the bound to a $90\%$
CL, since the experimental bounds are quoted with a scale factor of
$1.7$, and $\delta\text{Br}^{\text{Exp}}$ can be read from Eqs.~(\ref{eq:ZtollExp_a}-\ref{eq:ZtollExp_c}).
For the first experimental bound, $\ell_{i}=e$, which means that
FV bounds can be extracted by setting $\ell_{j}=\mu,\tau$. These
are given by $\sqrt{g_{L}^{2}|g_{L}^{\mu e}|^{2}+g_{R}^{2}|g_{R}^{\mu e}|^{2}}<8.43\times10^{-2}$
and $\sqrt{g_{L}^{2}|g_{L}^{\tau e}|^{2}+g_{R}^{2}|g_{R}^{\tau e}|^{2}}<8.43\times10^{-2}$,
respectively. Similarly, for the second bound we have $\ell_{i}=\mu$,
which means that we can obtain FV bounds by setting $\ell_{j}=e,\tau$.
The corresponding bounds are $\sqrt{g_{L}^{2}|g_{L}^{\mu e}|^{2}+g_{R}^{2}|g_{R}^{\mu e}|^{2}}<0.11$
and $\sqrt{g_{L}^{2}|g_{L}^{\tau\mu}|^{2}+g_{R}^{2}|g_{R}^{\tau\mu}|^{2}}<0.11$.
Finally, setting $\ell_{i}=\tau$, the FV bounds are obtained with
$\ell_{j}=\mu,e$, which read $\sqrt{g_{L}^{2}|g_{L}^{\tau e}|^{2}+g_{R}^{2}|g_{R}^{\tau e}|^{2}}<0.12$
and $\sqrt{g_{L}^{2}|g_{L}^{\tau\mu}|^{2}+g_{R}^{2}|g_{R}^{\tau\mu}|^{2}}<0.12$.

Notice that the bounds arising from $Z\to e^{+}e^{-}$ are stronger
than these arising from $Z\to\mu^{+}\mu^{-}$, which are in turn stronger
than these arising from $Z\to\tau^{+}\tau^{-}$, which is consistent
with the experimental limits given in Eqs. (\ref{eq:ZtollExp_a})
- (\ref{eq:ZtollExp_c}). However, these bounds are rather weak, which
implies that this is not a promising channel to search for FV in the
$Z$ sector.

\subsection{The $\ell_{i}\to\ell_{j}+\text{inv.}$ Decay}

\begin{figure}[t!]
\includegraphics[width=0.5\textwidth]{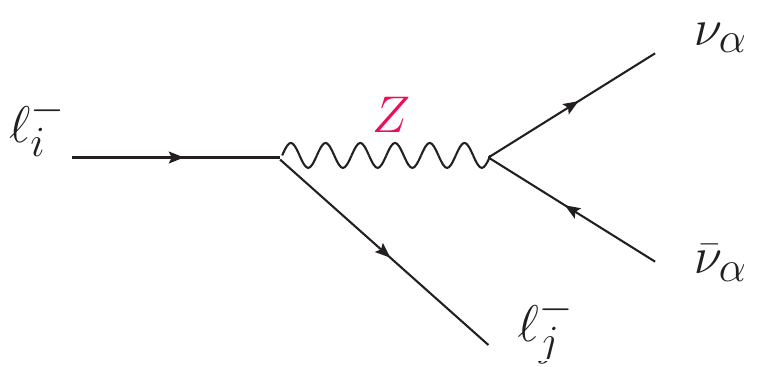} \caption{The decay $\ell_{i}\to\ell_{j}+\text{inv.}$ through a FV $Z$. Here,
we sum over all neutrino flavors $\alpha=e,\mu,\tau$.}
\label{fig4} 
\end{figure}

One of the potentially novel routes for probing the FV $Z$ interactions
might be obtained from the searches $\tau\to\mu\bar{\nu}\nu$, $\tau\to e\bar{\nu}\nu$,
and $\mu\to e\bar{\nu}\nu$, which have the signature $\ell_{i}\to\ell_{j}+\slashed{E}$.
A FV $Z$ can contribute to these processes through the diagram shown
in Figure~\ref{fig4}. In calculating this decay, we have to sum
over the three neutrino flavors, which are all indistinguishable and
appear as missing energy. As $m_{\tau}\gg m_{\mu}\gg m_{e}$, we can
treat the final state particles including $\ell_{j}$ as massless.
In this approximation, the decay width is given by

\begin{align}
\Gamma(\ell_{i}\to\ell_{j}+\text{inv.}) & \simeq\frac{g^{2}m_{i}}{3072\pi^{3}\cos^{2}\theta_{W}}\Big(|g_{L}^{ij}|^{2}+|g_{R}^{ij}|^{2}\Big)\Big[\Big(\frac{6-3\lambda-\lambda^{2}}{\lambda}\Big)+6\frac{(1-\lambda)}{\lambda^{2}}\log(1-\lambda)\Big],\label{eq:Zinv_width}\\
 & \simeq\frac{g^{2}m_{i}^{5}}{6144\pi^{3}\cos^{2}\theta_{W}m_{Z}^{4}}\Big(|g_{L}^{ij}|^{2}+|g_{R}^{ij}|^{2}\Big),
\end{align}
where $\lambda=\frac{m_{i}^{2}}{m_{Z}^{2}}$. The TWIST collaboration
reported bounds on $\mu^{+}\to e^{+}X^{0}$~\cite{TWIST:2014ymv},
whereas the Belle experiment reported bounds on $\tau\to\mu X^{0},eX^{0}$~\cite{Belle:2023ziz,Belle-II:2022heu},
with $X^{0}$ being an invisible vector boson. Unfortunately, these
experiments were designed for 2-body final states, where a mono-energetic
lepton is searched for. Thus, they are not applicable directly to
our 3-body decay shown in Figure~\ref{fig4}. However, FV $Z$ couplings
from such a decay can be constrained via the Michel parameters and
Fermi constant, which are measured by these experiments. We discuss
this in the next two sections.

\subsection{The Michel Parameters}

The Michel parameters describe the energy spectrum and angular distribution
of the charged lepton in leptonic decays like $\mu^{-}\to e^{-}\nu_{\mu}\bar{\nu}_{e}$~\cite{Michel:1949qe,Bouchiat:1957zz,Fetscher:1986uj}.
They are a set of measurable constants that encode the Lorentz structure
(scalar, vector and/or tensor) and the chiral structure of the weak
interaction, providing a powerful test of the SM's $V-A$ coupling.

The Lagrangian in Eq.~(\ref{eq:FV_lepton_Lag}) leads to the FV $Z$
mediating decay channels $\ell_{i}\to\ell_{j}\bar{\nu}_{\alpha}\nu_{\alpha}$
with the amplitude 
\begin{equation}
\mathcal{M}_{\alpha}=\frac{4G_{F}}{\sqrt{2}}[\bar{\ell}_{j}\gamma^{\mu}(r_{L}^{ij}P_{L}+r_{R}^{ij}P_{R})\ell_{i}][\bar{\nu}_{\alpha L}\gamma_{\mu}\nu_{\alpha L}],
\end{equation}
with $r_{L,R}^{ij}=\frac{\cos\theta_{W}}{g}g_{L,R}^{ij}$, where $g$
is the $SU(2)_{L}$ coupling and $\theta_{W}$ is the Weinberg mixing
angle. The decay width receives contributions from the SM decay in
addition to the FV channels $\ell_{i}\to\ell_{j}\bar{\nu}_{\alpha}\nu_{\alpha}$.
Then, following the same steps in~\cite{Marquez:2022bpg}, one can
write 
\begin{equation}
\frac{d^{2}\Gamma_{i}}{dx\,d\cos\theta}\propto x^{2}\left[(3-2x)(1+n_{\nu}|r_{L}^{ij}|^{2}+n_{\nu}|r_{R}^{ij}|^{2})+P_{\mu}\cos\theta(1-2x)(1+n_{\nu}|r_{L}^{ij}|^{2}-n_{\nu}|r_{R}^{ij}|^{2})\right],
\end{equation}
where $n_{\nu}=3$ is the number of neutrino flavors. For the muon
decay, this result has to be normalized by the factor $1+3|r_{L}^{\mu e}|^{2}+3|r_{R}^{\mu e}|^{2}$,
which gives the Michel parameters as 
\begin{equation}
\rho=\delta=\frac{3}{4},\,\eta=0,\,\xi=\frac{1-3|r_{L}^{\mu e}|^{2}-3|r_{R}^{\mu e}|^{2}}{1+3|r_{L}^{\mu e}|^{2}+3|r_{R}^{\mu e}|^{2}},
\end{equation}
where the experimental measurement is given by $\xi_{\mu}=1.0009\pm0.0016$~\cite{ParticleDataGroup:2024cfk}.
This leads to $\sqrt{|g_{L}^{\mu e}|^{2}+|g_{R}^{\mu e}|^{2}}<7.99\times10^{-3}$.
This bound is less stringent than the ones obtained from the Fermi
Constant and the Muon Lifetime as we will see next.

For the tau decay, the $\xi$ parameter can be written in the limit
$m_{e}/m_{\tau}=m_{\mu}/m_{\tau}\approx0$, as 
\begin{equation}
\xi=\frac{1-3|r_{L}^{\tau e}|^{2}-3|r_{R}^{\tau e}|^{2}-3|r_{L}^{\tau\mu}|^{2}-3|r_{R}^{\tau\mu}|^{2}}{1+3|r_{L}^{\tau e}|^{2}+3|r_{R}^{\tau e}|^{2}+3|r_{L}^{\tau\mu}|^{2}+3|r_{R}^{\tau\mu}|^{2}},
\end{equation}
with the experimental measurement given by $\xi_{\tau}=0.994\pm0.008$~\cite{ParticleDataGroup:2024cfk}.
This leads to the bound $\sqrt{|g_{L}^{\tau\mu}|^{2}+|g_{R}^{\tau\mu}|^{2}+|g_{L}^{\tau e}|^{2}+|g_{R}^{\tau e}|^{2}}<6.523\times10^{-2}$.

\subsection{Constraints from the Fermi Constant and the Muon Lifetime}

The Fermi constant $G_{F}$ is a well-measured constant that is extracted
from the lifetime of the muon, which in the SM is given by~\cite{ParticleDataGroup:2024cfk}

\begin{align}
\frac{1}{\tau_{\mu}} & =\Gamma_{\mu}=\frac{G_{F}^{2}m_{\mu}^{5}}{192\pi^{3}}F(\rho)\Bigg(1+H_{1}(\rho)\frac{\widehat{\alpha}(m_{\mu})}{\pi}+H_{2}(\rho)\frac{\widehat{\alpha}^{2}(m_{\mu})}{\pi^{2}}+H_{3}(\rho)\frac{\widehat{\alpha}^{3}(m_{\mu})}{\pi^{3}}\Bigg),\label{eq:mu_lifetime1}\\
F(\rho) & =1-8\rho+8\rho^{3}-\rho^{4}-12\rho^{2}\log(\rho)=0.99981295,\\
H_{1} & =\frac{25}{8}-\frac{\pi^{2}}{2}-(9+4\pi^{2}+12\log({\rho}))\rho+16\pi^{2}\rho^{3/2}+\mathcal{O}(\rho^{2})=-1.80793,\\
H_{2} & =\frac{156815}{5184}-\frac{518}{81}\pi^{2}-\frac{895}{36}\zeta(3)+\frac{67}{720}\pi^{4}+\frac{53}{6}\pi^{2}\log{(2)}\nonumber \\
 & -(0.042\pm0.002)_{\text{had}}-\frac{5}{4}\pi^{2}\sqrt{\rho}+\mathcal{O}(\rho)=6.64,\\
\widehat{\alpha}(m_{\mu})^{-1} & =\alpha(m_{\mu})^{-1}+\frac{1}{3\pi}\log{(\rho)}=135.901
\end{align}
and $\rho=m_{e}^{2}/m_{\mu}^{2}$. $H_{1}$ and $H_{2}$ represent
the QED corrections and were calculated in~\cite{vanRitbergen:1999fi,Steinhauser:1999bx,Nir:1989rm},
whereas $H_{3}$ has been estimated in~\cite{Pak:2008qt,Ferroglia:1999tg,Fael:2020tow,Czakon:2021ybq}
to be $-15.3\pm2.3$. A FV $Z$ coupling to $\mu e$ will impact $G_{F}$
and the muon lifetime via the decay shown in Fig.~\ref{fig4} above,
where the muon lifetime becomes 
\begin{equation}
\tau_{\mu}=\frac{1}{\Gamma_{\mu}^{\text{tot}}}=\frac{1}{\Gamma_{\mu}^{\text{SM}}+\Gamma_{\mu}^{\text{FV}}}\simeq\frac{1}{\Gamma_{\mu}^{\text{SM}}}-\frac{\Gamma_{\mu}^{\text{FV}}}{(\Gamma_{\mu}^{\text{SM}})^{2}}+\cdots,\label{eq:mu_lifetime2}
\end{equation}
where $\Gamma_{\mu}^{\text{FV}}$ is given by Eq.~(\ref{eq:Zinv_width}),
and we have assumed that $\Gamma_{\mu}^{\text{FV}}\ll\Gamma_{\mu}^{\text{SM}}$.
Thus, we obtain a $90\%$ C.L. bound on the FV $Z$ coupling to $\mu e$
by demanding that 
\begin{equation}
\Bigg|\frac{\Gamma_{\mu}^{\text{FV}}}{(\Gamma_{\mu}^{\text{SM}})^{2}}\Bigg|<1.645\delta\tau_{\mu},\label{eq:mu_lifetime_bound}
\end{equation}
which, given that $\tau_{\mu}=2.1969811\pm0.0000022~\mu s$~\cite{ParticleDataGroup:2024cfk}
yields $\sqrt{|g_{L}^{\mu e}|^{2}+|g_{R}^{\mu e}|^{2}}<9.45\times10^{-4}$.

\subsection{Constraints from $\ell_{i}\to\ell_{j}\gamma$}

\begin{figure}[t!]
\begin{centering}
\includegraphics[width=0.7\textwidth]{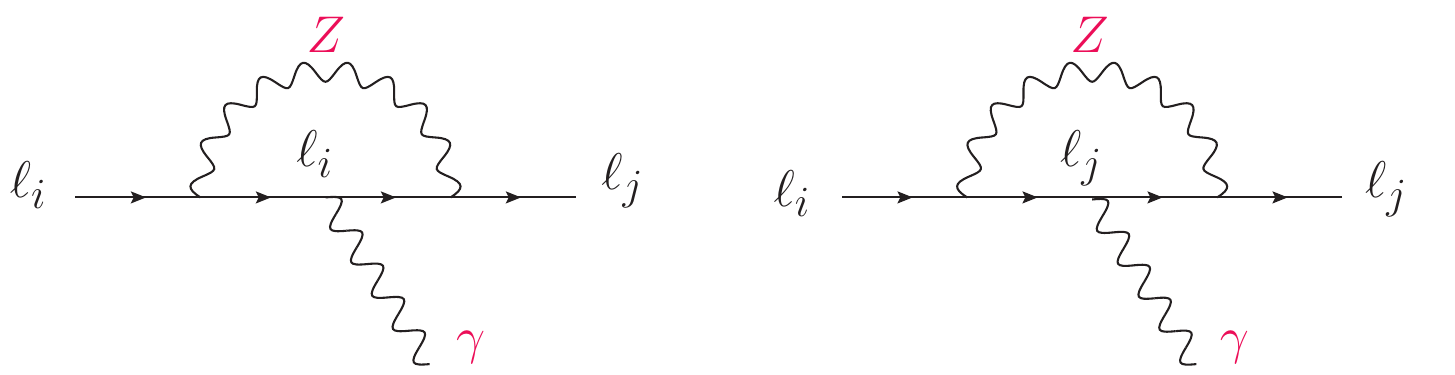} 
\par\end{centering}
\caption{The leading contributions to $\ell_{i}\to\ell_{j}\gamma$. Other contributions
where $\gamma$ is radiated from the initial or final states are subleading.}
\label{fig5} 
\end{figure}

A FV $Z$ can contribute to the FV decay $\ell_{i}\to\ell_{j}\gamma$
through the Feynman diagrams shown in Fig.~\ref{fig5}, where $\ell_{i}\neq\ell_{j}$.
In addition to these vertex corrections, there are contributions where
the photon is radiated from the initial and final state leptons, however,
these diagrams are subleading and we neglect them here. We relegate
the detailed calculation to Appendix~\ref{appendix_a}. Neglecting
terms of $\mathcal{O}(\frac{m_{i,j}^{2}}{m_{Z}^{2}})$, the decay
width is given by 
\begin{equation}
\Gamma(\ell_{i}\to\ell_{j}\gamma)\simeq\alpha m_{i}\Big(\frac{3}{32\pi^{2}}\Big)^{2}\Big(g_{L}^{2}|g_{L}^{ij}|^{2}+g_{R}^{2}|g_{R}^{ij}|^{2}\Big).\label{eq:l_gamma_width}
\end{equation}

The most recent bound on the $\tau\to\mu\gamma$ decay is given by
$\text{Br}(\tau\to\mu\gamma)<4.2\times10^{-8}$~\cite{BaBar:2009hkt}.
This translates to $\sqrt{g_{L}^{2}|g_{L}^{\tau\mu}|^{2}+g_{R}^{2}|g_{R}^{\tau\mu}|^{2}}<2.85\times10^{-7}$.
Similarly, the bounds on the $\tau\to e\gamma$ and $\mu\to e\gamma$
decays are given by $\text{Br}(\tau\to e\gamma)<3.3\times10^{-8}$~\cite{BaBar:2009hkt}
and $\text{Br}(\mu\to e\gamma)<1.5\times10^{-13}$~\cite{MEGII:2025gzr},
which lead to the FV bounds $\sqrt{g_{L}^{2}|g_{L}^{\tau e}|^{2}+g_{R}^{2}|g_{R}^{\tau e}|^{2}}<2.53\times10^{-7}$
and $\sqrt{g_{L}^{2}|g_{L}^{\mu e}|^{2}+g_{R}^{2}|g_{R}^{\mu e}|^{2}}<1.34\times10^{-12}$.
We can see that these bounds are quite stringent, making this channel
one of the best places to search for FV in the $Z$ sector.

The Belle II and MEG II experiments are expected to improve the bounds
by about an order of magnitude~\cite{Calibbi:2017uvl}. Specifically,
the projected Belle II bounds are expected to be $\text{Br}(\tau\to\mu\gamma)<1\times10^{-9}$,
$\text{Br}(\tau\to e\gamma)<5\times10^{-9}$, whereas the projected
MEG II bound is expected to be $\text{Br}(\mu\to e\gamma)<5\times10^{-14}$.
These would yield the bounds $\sqrt{g_{L}^{2}|g_{L}^{\tau\mu}|^{2}+g_{R}^{2}|g_{R}^{\tau\mu}|^{2}}<4.4\times10^{-8}$,
$\sqrt{g_{L}^{2}|g_{L}^{\tau e}|^{2}+g_{R}^{2}|g_{R}^{\tau e}|^{2}}<9.84\times10^{-8}$,
and $\sqrt{g_{L}^{2}|g_{L}^{\mu e}|^{2}+g_{R}^{2}|g_{R}^{\mu e}|^{2}}<4.63\times10^{-13}$,
respectively.

\subsection{Constraints from $\ell_{i}\to\ell_{j}+\ell_{k}+\bar{\ell}_{k}$}

\begin{figure}[t!]
\begin{centering}
\includegraphics[width=0.7\textwidth]{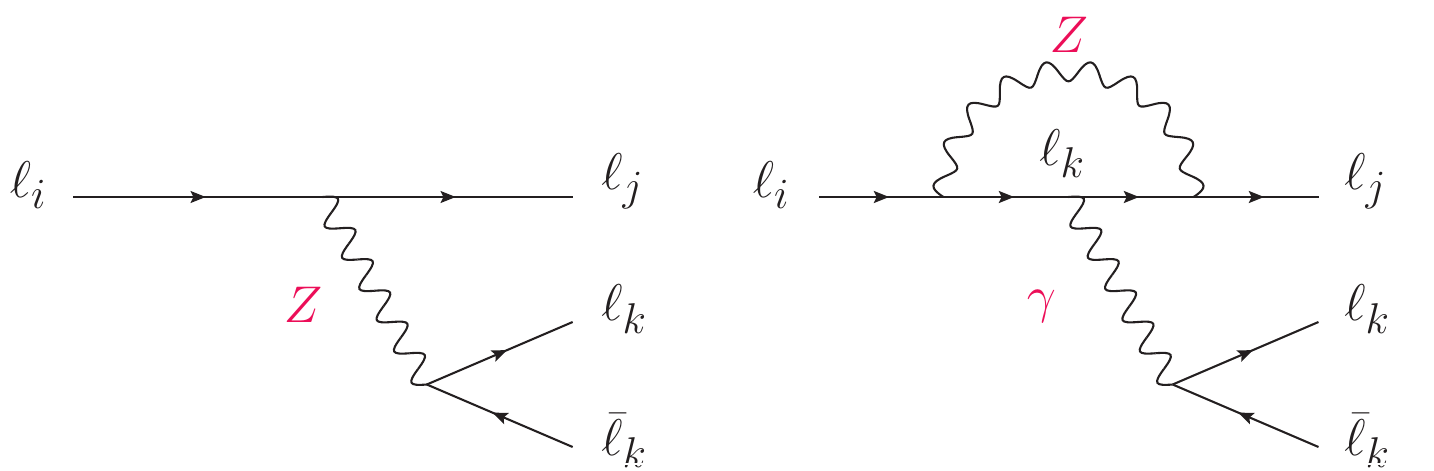} 
\par\end{centering}
\caption{The leading diagram for the process $\ell_{i}\to\ell_{k}+\bar{\ell}_{k}+\ell_{j}$.}
\label{fig6} 
\end{figure}

Stringent constraints on the FV $Z$ couplings can be obtained from
the decays $\ell_{i}\to\ell_{j}+\ell_{k}+\bar{\ell}_{k}$ where $\ell_{j}$
may or may not be identical to $\ell_{k}$. The leading contributions
to these processes are shown in Figure~\ref{fig6}.

Notice that for the diagram on the right, it is possible to integrate
out the loop and write it as an effective interaction at tree level.
This can be achieved by utilizing the results of $\ell_{i}\to\ell_{j}\gamma$
(see Eq.~(\ref{eq:l_gamma_mat_2})). We present the full calculation
in Appendix~\ref{appendix_b}. Explicit calculation shows that the
loop calculation on the right dominates the one on the left by orders
of magnitude in spite of the loop factor. The reason for this is the
heavy mass in the $Z$ propagator which suppresses the amplitude significantly
enough to compensate for the loop factor. Thus, we neglect the contribution
from this diagram. Assuming $m_{i}\gg m_{j},m_{k}$, which is a valid
assumption for the decays considered, we can treat the final state
particles as massless. We only keep the mass of $\ell_{k}$ to avoid
any IR divergence in the second diagram. With these approximations,
the decay width simplifies greatly and can be approximated as 
\begin{equation}
\Gamma(\ell_{i}\to\ell_{j}\ell_{k}\bar{\ell}_{k})\simeq\frac{3\alpha^{2}}{2048\pi^{5}S}\frac{m_{i}^{3}}{m_{k}^{2}}\Big(g_{L}^{2}|g_{L}^{ij}|^{2}+g_{R}^{2}|g_{R}^{ij}|^{2}\Big),\label{eq:lto3l_width}
\end{equation}
where $S=2$ is a possible symmetry factor if two of the final state
particles are identical, i.e., $j=k$. Now, we can utilize the experimental
bounds to set limits on the FV $Z$ couplings. The first bound we
consider is on the decay $\tau\to3\mu$. The experimental limit is
given by $\text{Br}(\tau\to3\mu)<2.1\times10^{-8}$~\cite{ParticleDataGroup:2024cfk},
which translates to $\sqrt{g_{L}^{2}|g_{L}^{\tau\mu}|^{2}+g_{R}^{2}|g_{R}^{\tau\mu}|^{2}}<8.64\times10^{-7}$.
Bounds on the same coupling combinations can be obtained from the
experimental limit $\text{Br}(\tau\to\mu^{-}e^{+}e^{-})<1.8\times10^{-8},$
and $\text{Br}(\tau\to\mu^{+}e^{-}e^{-})<1.5\times10^{-8}$~\cite{ParticleDataGroup:2024cfk},
which yield $\sqrt{g_{L}^{2}|g_{L}^{\tau\mu}|^{2}+g_{R}^{2}|g_{R}^{\tau\mu}|^{2}}<5.66\times10^{-7}$
and $\sqrt{g_{L}^{2}|g_{L}^{\tau\mu}|^{2}+g_{R}^{2}|g_{R}^{\tau\mu}|^{2}}<7.3\times10^{-7}$,
respectively. To obtain bounds on the $\tau e$ couplings, we use
the experimental bounds on $\text{Br}(\tau\to3e)<2.7\times10^{-8}$,
$\text{Br}(\tau\to\mu^{+}\mu^{-}e^{-})<2.7\times10^{-8}$ and $\text{Br}(\tau\to\mu^{-}\mu^{-}e^{+})<1.7\times10^{-8}$~\cite{ParticleDataGroup:2024cfk},
which yield the bounds $\sqrt{g_{L}^{2}|g_{L}^{\tau e}|^{2}+g_{R}^{2}|g_{R}^{\tau e}|^{2}}<4.72\times10^{-9}$,
$\sqrt{g_{L}^{2}|g_{L}^{\tau e}|^{2}+g_{R}^{2}|g_{R}^{\tau e}|^{2}}<3.34\times10^{-9}$
and $\sqrt{g_{L}^{2}|g_{L}^{\tau e}|^{2}+g_{R}^{2}|g_{R}^{\tau e}|^{2}}<3.75\times10^{-9}$,
respectively. Finally, the experimental measurement of $\text{Br}(\mu\to3e)<1\times10^{-12}$~\cite{ParticleDataGroup:2024cfk}
gives the bound $\sqrt{g_{L}^{2}|g_{L}^{\mu e}|^{2}+g_{R}^{2}|g_{R}^{\mu e}|^{2}}<7.17\times10^{-13}$.
We can see that for each combination of couplings, bounds arising
from different searches are close to one another, however, the bounds
arising from mixed final states are slightly stronger than those obtained
from identical final states.

In general, $\ell_{i}\to 3\ell_{j}$ decays are one of the most promising
channels for searching for FV, with future experiments planned to
take place. The Belle II and Mu3e experiments are expected to provide
stronger bounds on these decays~\cite{Aushev:2010bq,Belle-II:2018jsg,Calibbi:2017uvl, Perrevoort:2018ttp}
(see~\cite{Banerjee:2022vdd} also for clean projections). In particular,
the projected Belle II bounds on $\text{Br}(\tau\to3\mu)$ and $\text{Br}(\tau\to3e)$
are expected to be $5\times10^{-10}$, which translates into $\sqrt{g_{L}^{2}|g_{L}^{\tau\mu}|^{2}+g_{R}^{2}|g_{R}^{\tau\mu}|^{2}}<1.33\times10^{-7}$
and $\sqrt{g_{L}^{2}|g_{L}^{\tau e}|^{2}+g_{R}^{2}|g_{R}^{\tau e}|^{2}}<6.43\times10^{-10}$
respectively. Similarly, the projected bound on $\text{Br}(\mu\to3e)$
from the Mu3e experiment is expected to reach $1\times10^{-16}$,
which becomes $\sqrt{g_{L}^{2}|g_{L}^{\mu e}|^{2}+g_{R}^{2}|g_{R}^{\mu e}|^{2}}<7.17\times10^{-15}$.

\subsection{Constraints from FV Meson Decays}

\begin{figure}[t!]
\begin{centering}
\includegraphics[width=0.5\textwidth]{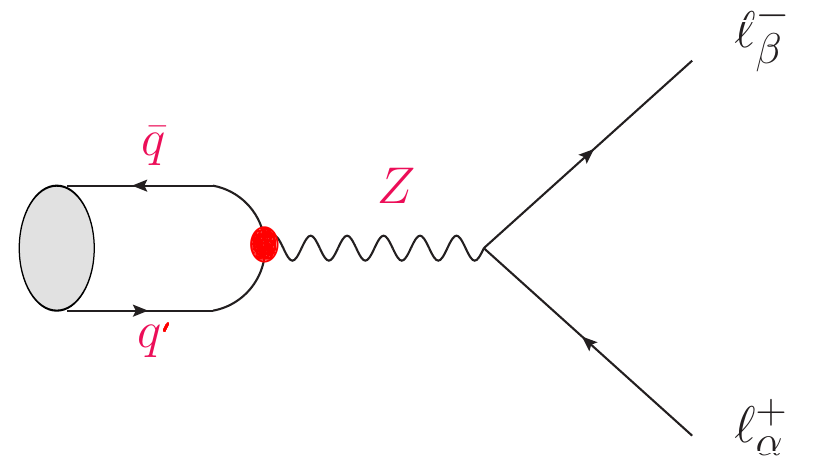} 
\par\end{centering}
\caption{FV meson decays mediated by the $Z$ boson. The vertex in red is a
tree-level vertex for vector mesons and for the pseudoscalar mesons
$\eta$, $\eta^{\prime}$ and $\pi^{0}$, whereas for the pseudoscalar
mesons $B_{d}$, $B_{s}$, $D^{0}$, $K_{L}^{0}$ and $K_{S}^{0}$,
it represents an effective coupling that arises through a one-loop
triangle diagram mediated by the W bosons and quarks, where the FV
is induced via the off-diagonal CKM matrix elements.}
\label{fig7} 
\end{figure}

It was suggested in~\cite{Hazard:2017swy} (see also~\cite{Delepine:2015mca})
that experimental bounds on FV meson decays $M\to\ell_{i}\ell_{j}$
can be used to set bounds on the FV Wilson coefficients in an EFT.
Such decays can also be used to constrain the FV $Z$ couplings directly
by considering the diagram in Figure~\ref{fig7}. As this decay proceeds
via a vector current mediated by the $Z$ boson, we can utilize the
technique of current algebra, and write the interaction of the meson
with the $Z$ boson as 
\begin{equation}
\mathcal{L}_{\text{int}}=\frac{e}{\sin{\theta_{W}}}Z_{\mu}J_{Z}^{\mu},\label{eq:Current_int_1}
\end{equation}
where 
\begin{equation}
J_{Z}^{\mu}=\sum_{i}\Big[\frac{1}{\cos\theta_{W}}\bar{\psi}_{i}\gamma^{\mu}T^{3}\psi_{i}-\frac{\sin^{2}\theta_{W}}{\cos\theta_{W}}Q_{i}\bar{\psi}_{i}\gamma^{\mu}\psi_{i}\Big],\label{eq:Current_int_2}
\end{equation}
and the sum is over all quarks. Thus, in the language of the current
algebra, we can simply express the matrix element as follows 
\begin{equation}
\mathcal{M}=\frac{ie}{\sin{\theta_{W}}}\bra{0}J_{Z}^{\mu}\ket{M(p)}\Bigg(\frac{g_{\mu\nu}-\frac{1}{m_{Z}^{2}}p_{\mu}p_{\nu}}{p^{2}-m_{Z}^{2}}\Bigg)\bar{u}(k_{1})\gamma^{\nu}(g_{L}^{ij*}P_{L}+g_{R}^{ij*}P_{R})v(k_{2}),\label{eq:meson_decay_mat1}
\end{equation}
where $p$ is the momentum of the meson. As we can see, the neutral
vector current $J_{Z}^{\mu}$ allows the meson to turn into a $Z$
boson, which in turn can decay via a FV interaction to leptons of
different flavors. In general, the QCD matrix element is measured
experimentally in terms of the decay constant and it depends on the
type of mesons. For vector mesons, which are mesons with quantum numbers
$J^{pc}=1^{--}$, the QCD matrix element can be expressed as 
\begin{equation}
\bra{0}J_{Z}^{\mu}\ket{V(q)}=f_{V}m_{V}\epsilon^{\mu}(q),\label{eq:vec_meson_mat}
\end{equation}
where $f_{V}$ is the decay constant and $m_{V}$ is the mass of the
vector meson. This can be used in Eq.~(\ref{eq:meson_decay_mat1})
to find the decay width. Assuming without loss of generality that
$m_{i}>m_{j}$ and dropping $m_{j}$ since $m_{\tau}\gg m_{\mu}\gg m_{e}$,
we find 
\begin{equation}
\Gamma(V\to\bar{\ell}_{i}\ell_{j})\simeq\frac{e^{2}f_{V}^{2}m_{V}^{3}}{48\pi\sin^{2}\theta_{W}m_{Z}^{4}}\Big(|g_{L}^{ij}|^{2}+|g_{R}^{ij}|^{2}\Big)\Bigg[1-\frac{m_{V}^{2}}{m_{Z}^{2}}\Bigg]^{-2}\Bigg[1-\frac{m_{i}^{2}}{m_{V}^{2}}\Bigg]^{2}\Bigg[2+\frac{m_{i}^{2}}{m_{V}^{2}}\Bigg].\label{eq:vec_meson_width}
\end{equation}

The vector mesons for which experimental bounds on leptonic FV decays
exist are: $\Upsilon(1S)$, $\Upsilon(2S)$, $\Upsilon(3S)$, $J/\psi$
and $\phi$. The relevant information of these mesons, including their
masses, decay constant and decay widths, are provided in Table~\ref{tab4}.
The latest bound on the FV decays of all of these mesons can be found
in~\cite{ParticleDataGroup:2024cfk}. For $\Upsilon(1S)$, the bound
is given by $\text{Br}(\Upsilon(1S)\to\tau\mu)<6\times10^{-6}$, which
translates into the bound $\sqrt{|g_{L}^{\tau\mu}|^{2}+|g_{R}^{\tau\mu}|^{2}}<0.11$.
Similarly, the FV limits from $\Upsilon(2S)$ are given by $\text{Br}(\Upsilon(2S)\to\tau\mu)<3.3\times10^{-6}$
and $\text{Br}(\Upsilon(2S)\to\tau e)<3.2\times10^{-6}$, which translate
to $\sqrt{|g_{L}^{\tau\mu}|^{2}+|g_{R}^{\tau\mu}|^{2}}<7.81\times10^{-2}$
and $\sqrt{|g_{L}^{\tau e}|^{2}+|g_{R}^{\tau e}|^{2}}<7.69\times10^{-2}$,
respectively. The experimental bounds from $\Upsilon(3S)$ read $\text{Br}(\Upsilon(3S)\to\tau\mu)<3.1\times10^{-6}$
and $\text{Br}(\Upsilon(3S)\to\tau e)<4.2\times10^{-6}$ from which
we obtain $\sqrt{|g_{L}^{\tau\mu}|^{2}+|g_{R}^{\tau\mu}|^{2}}<5.11\times10^{-2}$
and $\sqrt{|g_{L}^{\tau e}|^{2}+|g_{R}^{\tau e}|^{2}}<5.95\times10^{-2}$,
respectively. The experimental measurements of the FV decay of $J/\psi$
are given by $\text{Br}(J/\psi\to\tau\mu)<2\times10^{-6}$, $\text{Br}(J/\psi\to\tau e)<8.3\times10^{-6}$
and $\text{Br}(J/\psi\to\mu e)<1.6\times10^{-7}$, from which we obtain
$\sqrt{|g_{L}^{\tau\mu}|^{2}+|g_{R}^{\tau\mu}|^{2}}<0.95$, $\sqrt{|g_{L}^{\tau e}|^{2}+|g_{R}^{\tau e}|^{2}}<1.93$
and $\sqrt{|g_{L}^{\mu e}|^{2}+|g_{R}^{\mu e}|^{2}}<0.19$, respectively.
Finally, the bound on the decay of $\phi$ is given by $\text{Br}(\phi\to\mu e)<2\times10^{-6}$,
which translates to the rather weak bound $\sqrt{|g_{L}^{\mu e}|^{2}+|g_{R}^{\mu e}|^{2}}<42.93$.
The last bound seems to indicate that perturbativity is lost, however,
the experimental limits are simply not strong enough to yield a meaningful
bound. 
\begin{table*}[t!]
\centering{}%
\begin{tabular}{|c|c|c|c|}
\hline 
\textbf{Meson} & \textbf{$m_{V}$} & \textbf{$\Gamma_{M_{V}}$} & \textbf{$f_{V}$} \tabularnewline
\hline 
$\Upsilon(1S)$ & $9.4603$ & $5.402\times10^{-5}$ & $0.649$ \tabularnewline
$\Upsilon(2S)$ & $10.02326$ & $3.198\times10^{-5}$ & $0.481$ \tabularnewline
$\Upsilon(3S)$ & $10.3552$ & $2.032\times10^{-5}$ & $0.539$ \tabularnewline
$J/\psi$ & $3.0969$ & $9.29\times10^{-5}$ & $0.418$ \tabularnewline
$\phi$ & $1.019461$ & $4.249\times10^{-3}$ & $0.241$\tabularnewline
$\rho$ & $0.77526$ & $0.1478$ & $0.2094$\tabularnewline
$\omega$ & $0.78266$ & $8.68\times10^{-3}$ & $0.2094$\tabularnewline
\hline 
\end{tabular}\caption{Mass, decay width and decay constant of the relevant vector mesons~\cite{ParticleDataGroup:2024cfk}.
All values are in GeV.}
\label{tab4} 
\end{table*}

Next, we turn our attention to pseudoscalar mesons~\footnote{The only scalar meson that has FV bounds is the $f^{0}$ meson, however,
we were unable to find a measured value of its decay constant.}, which are mesons with quantum numbers $J^{pc}=0^{-+}$. Table~\ref{tab5}
contains all the relevant information for these mesons. The QCD matrix
element can be expressed as 
\begin{equation}
\bra{0}J_{Z}^{\mu}\ket{P(p)}=-if_{P}p^{\mu},\label{eq:PS_meson_mat}
\end{equation}
which when used in Eq.~(\ref{eq:meson_decay_mat1}) yields the decay
width 
\begin{equation}
\Gamma(P\to\ell_{i}\ell_{j})\simeq\frac{ne^{2}f_{P}^{2}}{16\pi\sin^{2}\theta_{W}}\frac{m_{P}m_{i}^{2}}{m_{Z}^{4}}\Big(|g_{L}^{ij}|^{2}+|g_{R}^{ij}|^{2}\Big)\Bigg[1-\frac{m_{i}^{2}}{m_{P}^{2}}\Bigg]^{2},\label{eq:PS_width}
\end{equation}
where $n=2$ for $P=\pi^{0},\eta,\eta'$ and $n=1$ for all other
mesons, since for the former mesons, experimental bounds are quoted
on $\ell_{i}^{\pm}\ell_{j}^{\mp}+\ell_{i}^{\mp}\ell_{j}^{\pm}$, and
thus there is an additional factor of 2 arising from the h.c. part
of the Lagrangian. The experimental bounds on the FV decays of $\eta$,
$\eta'$ and $\pi^{0}$ are given by $\text{Br}(\eta\to\mu e)<6\times10^{-6}$,
$\text{Br}(\eta'\to\mu e)<4.7\times10^{-4}$ and $\text{Br}(\pi^{0}\to\mu e)<3.6\times10^{-6}$,
respectively~\cite{ParticleDataGroup:2024cfk}. They translate to
the following bounds $\sqrt{|g_{L}^{\mu e}|^{2}+|g_{R}^{\mu e}|^{2}}<22.72$,
$\sqrt{|g_{L}^{\mu e}|^{2}+|g_{R}^{\mu e}|^{2}}<2.16\times10^{3}$
and $\sqrt{|g_{L}^{\mu e}|^{2}+|g_{R}^{\mu e}|^{2}}<5.7\times10^{-2}$,
respectively, and we can see that the experimental bounds from the
decay of $\eta$ and $\eta'$ are too weak to be meaningful, unlike
the bounds from the decay of $\pi^{0}$.

Next, we consider the FV decays from $B$ mesons. The limits on the
FV decays of $B_{d}$ are given by~\cite{ParticleDataGroup:2024cfk}
$\text{Br}(B_{d}\to\tau\mu)<1.4\times10^{-5}$, $\text{Br}(B_{d}\to\tau e)<1.6\times10^{-5}$
and $\text{Br}(B_{d}\to\mu e)<1\times10^{-9}$, which translate to
$\sqrt{|g_{L}^{\tau\mu}|^{2}+|g_{R}^{\tau\mu}|^{2}}<3.42\times10^{-4}$,
$\sqrt{|g_{L}^{\tau e}|^{2}+|g_{R}^{\tau e}|^{2}}<3.66\times10^{-4}$
and $\sqrt{|g_{L}^{\mu e}|^{2}+|g_{R}^{\mu e}|^{2}}<4.3\times10^{-5}$.
Similarly, the experimental limits on the FV decays of $B_{s}$ are
given by $\text{Br}(B_{s}\to\tau\mu)<4.2\times10^{-4}$, $\text{Br}(B_{s}\to\tau e)<1.4\times10^{-3}$
and $\text{Br}(B_{s}\to\mu e)<5.4\times10^{-9}$~\cite{ParticleDataGroup:2024cfk},
which yield the bounds $\sqrt{|g_{L}^{\tau\mu}|^{2}+|g_{R}^{\tau\mu}|^{2}}<7.24\times10^{-3}$,
$\sqrt{|g_{L}^{\tau e}|^{2}+|g_{R}^{\tau e}|^{2}}<4.18\times10^{-2}$
and $\sqrt{|g_{L}^{\mu e}|^{2}+|g_{R}^{\mu e}|^{2}}<8.21\times10^{-5}$,
respectively.

Finally, we consider the FV decays of the neutral mesons $D^{0}$
and $K_{L}^{0}$. The experimental limits are given by $\text{Br}(D^{0}\to\mu e)<1.3\times10^{-8}$
and $\text{Br}(K_{L}^{0}\to\mu e)<4.7\times10^{-12}$, which translate
to $\sqrt{|g_{L}^{\mu e}|^{2}+|g_{R}^{\mu e}|^{2}}<4.51\times10^{-4}$
and $\sqrt{|g_{L}^{\mu e}|^{2}+|g_{R}^{\mu e}|^{2}}<6.57\times10^{-8}$
respectively. 
\begin{table*}[t!]
\centering{}%
\begin{tabular}{|c|c|c|c|}
\hline 
\textbf{Meson} & \textbf{$m_{P}$} & \textbf{$\Gamma_{M_{P}}$} & \textbf{$f_{P}$} \tabularnewline
\hline 
$\eta$ & $0.547862$ & $1.31\times10^{-6}$ & $0.108$ \tabularnewline
$\eta'$ & $0.95778$ & $1.88\times10^{-4}$ & $0.089$ \tabularnewline
$\pi^{0}$ & $0.1349768$ & $7.80421\times10^{-9}$ & $0.13041$ \tabularnewline
$B_{d}$ & $5.27972$ & $4.33681\times10^{-13}$ & $0.186$ \tabularnewline
$B_{s}$ & $5.36693$ & $4.30841\times10^{-13}$ & $0.224$ \tabularnewline
$D^{0}$ & $1.86484$ & $1.60345\times10^{-12}$ & $0.2074$ \tabularnewline
$K_{L}^{0}$ & $0.497611$ & $1.28596\times10^{-17}$ & $0.155$ \tabularnewline
$K_{S}^{0}$ & $0.497611$ & $7.3475\times10^{-15}$ & $0.155$ \tabularnewline
\hline 
\end{tabular}\caption{Mass, decay width and decay constant of the relevant pseudo-scalar
mesons~\cite{ParticleDataGroup:2024cfk}. All values are in GeV.}
\label{tab5} 
\end{table*}

We can see from the above constraints, that meson decays can provide
relatively strong constraints that are more stringent than direct
searches, however, they fall short of competing with bounds obtained
from the usual channels, like $\ell_{i}\to\ell_{j}\gamma$ and $\ell_{i}\to3\ell_{j}$.
Nonetheless, searches for FV meson decays remain promising as potential
channels to search for FV in the $Z$ sector, and thus merit further
interest.

\subsection{$\tau\to\mu(e)+\text{Meson}$ Decays}

The same treatment introduced in the previous section can be used
to set bounds on FV $\tau$ decays to a muon or an electron accompanied
by a meson. Such decays proceed via a diagram like Figure~\ref{fig7}
in reverse, with the initial state being $\ell_{i}=\tau$, and the
final state being $\ell_{j}=\mu,e$ + meson. The same current algebra
technique can be used, and the results are quite similar. In our calculation
we drop $m_{j}$ as usual. The decay width involving a vector meson
is given by 
\begin{equation}
\Gamma(\tau\to\ell V)\simeq\frac{e^{2}f_{V}^{2}m_{\tau}^{3}}{32\pi\sin^{2}\theta_{W}m_{Z}^{4}}\Big(|g_{L}^{\tau\ell}|^{2}+|g_{R}^{\tau\ell}|^{2}\Big)\Bigg[1-\frac{m_{V}^{2}}{m_{\tau}^{2}}\Bigg]^{2}\Bigg[1+\frac{2m_{V}^{2}}{m_{\tau}^{2}}\Bigg]\Bigg[1-\frac{m_{V}^{2}}{m_{Z}^{2}}\Bigg]^{-2}.
\end{equation}

The latest experimental measurements of the FV $\tau$ decays to $\mu,e$
+ meson can be found in~\cite{ParticleDataGroup:2024cfk}. For decays
involving the $\rho^{0}$ meson, the experimental measurements are
given by $\text{Br}(\tau\to\mu\rho^{0})<1.7\times10^{-8}$ and $\text{Br}(\tau\to e\rho^{0})<2.2\times10^{-8}$,
which translate to the limits $\sqrt{|g_{L}^{\tau\mu}|^{2}+|g_{R}^{\tau\mu}|^{2}}<5.5\times10^{-5}$
and $\sqrt{|g_{L}^{\tau e}|^{2}+|g_{R}^{\tau e}|^{2}}<6.26\times10^{-5}$,
respectively. The experimental measurements involving decays to the
$\omega$ meson read $\text{Br}(\tau\to\mu\omega)<3.9\times10^{-8}$
and $\text{Br}(\tau\to e\omega)<2.4\times10^{-8}$, which yield the
bounds $\sqrt{|g_{L}^{\tau\mu}|^{2}+|g_{R}^{\tau\mu}|^{2}}<8.34\times10^{-5}$
and $\sqrt{|g_{L}^{\tau e}|^{2}+|g_{R}^{\tau e}|^{2}}<6.55\times10^{-5}$.
Finally, the experimental limits on the decays involving the $\phi$
meson are given by $\text{Br}(\tau\to\mu\phi)<2.3\times10^{-8}$ and
$\text{Br}(\tau\to e\phi)<2\times10^{-8}$, which translate to $\sqrt{|g_{L}^{\tau\mu}|^{2}+|g_{R}^{\tau\mu}|^{2}}<6.12\times10^{-5}$
and $\sqrt{|g_{L}^{\tau e}|^{2}+|g_{R}^{\tau e}|^{2}}<5.71\times10^{-5}$,
respectively.

Next, we turn to $\tau$ decays involving pseudoscalar mesons. The
decay width of these processes is given by 
\begin{equation}
\Gamma(\tau\to\ell P)\simeq\frac{e^{2}f_{P}^{2}m_{\tau}^{3}}{32\pi\sin^{2}\theta_{W}m_{Z}^{4}}\Big(|g_{L}^{\tau\ell}|^{2}+|g_{R}^{\tau\ell}|^{2}\Big)\Bigg[1-\frac{m_{P}^{2}}{m_{\tau}^{2}}\Bigg]^{2}.
\end{equation}

Here, the FV $\tau$ decays involve $\pi^{0}$, $K_{S}^{0}$, $\eta$
and $\eta^{\prime}$. The bounds on the decays that involve $\pi^{0}$
are given by $\text{Br}(\tau\to\mu\pi^{0})<1.1\times10^{-7}$ and
$\text{Br}(\tau\to e\pi^{0})<1\times10^{-8}$, which translate to
$\sqrt{|g_{L}^{\tau\mu}|^{2}+|g_{R}^{\tau\mu}|^{2}}<2.15\times10^{-4}$
and $\sqrt{|g_{L}^{\tau e}|^{2}+|g_{R}^{\tau e}|^{2}}<1.83\times10^{-4}$.
The bounds involving $K_{S}^{0}$ are given by $\text{Br}(\tau\to\mu K_{S}^{0})<2.3\times10^{-8}$
and $\text{Br}(\tau\to eK_{S}^{0})<2.6\times10^{-8}$, and these translate
to $\sqrt{|g_{L}^{\tau\mu}|^{2}+|g_{R}^{\tau\mu}|^{2}}<8.92\times10^{-5}$
and $\sqrt{|g_{L}^{\tau e}|^{2}+|g_{R}^{\tau e}|^{2}}<9.48\times10^{-5}$.
The bounds involving $\eta$ are given by $\text{Br}(\tau\to\mu\eta)<6.5\times10^{-8}$
and $\text{Br}(\tau\to e\eta)<9.8\times10^{-8}$, from which we obtain
$\sqrt{|g_{L}^{\tau\mu}|^{2}+|g_{R}^{\tau\mu}|^{2}}<2.19\times10^{-4}$
and $\sqrt{|g_{L}^{\tau e}|^{2}+|g_{R}^{\tau e}|^{2}}<2.61\times10^{-4}$.
Finally, the bounds involving $\eta^{\prime}$ are given by $\text{Br}(\tau\to\mu\eta^{\prime})<1.3\times10^{-7}$
and $\text{Br}(\tau\to e\eta^{\prime})<1.6\times10^{-7}$, from which
we obtain $\sqrt{|g_{L}^{\tau\mu}|^{2}+|g_{R}^{\tau\mu}|^{2}}<4.8\times10^{-4}$
and $\sqrt{|g_{L}^{\tau e}|^{2}+|g_{R}^{\tau e}|^{2}}<5.32\times10^{-4}$.
The Belle II experiment is expected to improve the bound on the $\tau\to\mu\eta$
decay to $10^{-9}$~\cite{Belle-II:2010dht}, which would translate
to $\sqrt{|g_{L}^{\tau\mu}|^{2}+|g_{R}^{\tau\mu}|^{2}}<2.72\times10^{-5}$.

Inspecting the bounds, we can see they are in general stronger than
those obtained from meson decays by roughly one to two orders of magnitude.
Thus, the FV $\tau$ decays involving mesonic final states are also
a promising channel for probing FV, which highlights the potential
of $\tau$ factories. We conclude this section by pointing out that
the same treatment cannot be applied to the $\mu$ lepton since it
is lighter than all other mesons, and thus cannot decay to a final
state involving any meson.

\subsection{$\mu$ to $e$ Conversion in Nuclei}

\begin{figure}[t!]
\begin{centering}
\includegraphics[width=0.9\textwidth]{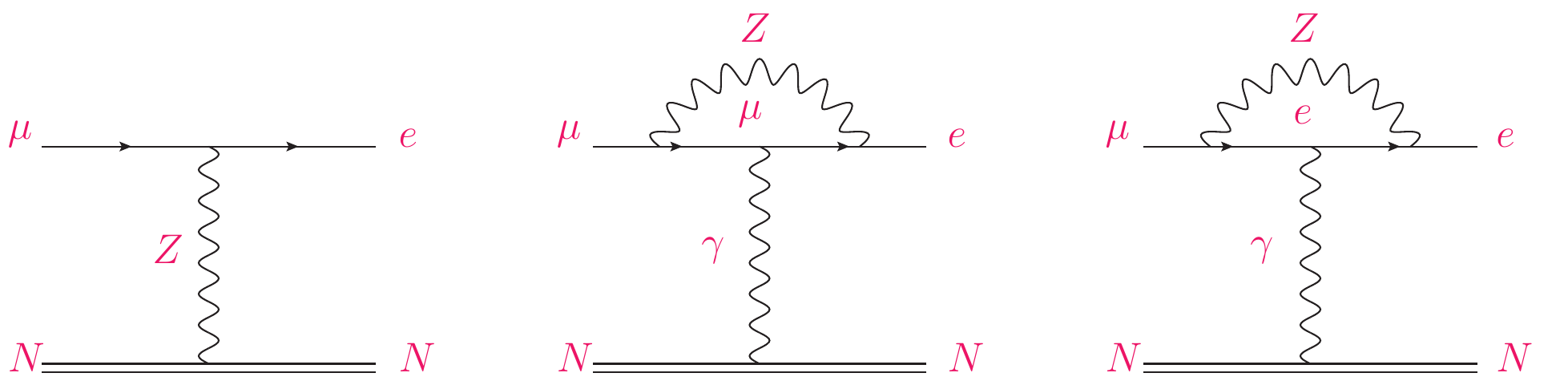} 
\par\end{centering}
\caption{The muon conversion in nuclei at tree level and at one loop.}
\label{fig8} 
\end{figure}

Very stringent constraints on the FV $Z$ couplings to $\mu e$ can
be obtained from the experimental searches for muon conversion in
nuclei. Such a conversion can proceed via the FV $Z$ already at tree
level as in the left diagram of Figure~\ref{fig8}. It can also proceed
at one loop as in the remaining two diagrams in Figure~\ref{fig8}.
Notice that in the loop diagrams, the $Z$ loop can be integrated
out to become an effective vertex by utilizing the results of $\ell_{i}\to\ell_{j}\gamma$
decay. Here we highlight the overall results and present the full
details in Appendix~\ref{appendix_c}.

The experimental limits on the conversion rate of $\mu\to e$ in nuclei
are quoted as the branching fraction of the conversion rate relative
to the muon capture rate by the target nucleus. According to the SINDRUM
II collaboration~\cite{SINDRUMII:2006dvw}, gold yields the strongest
bounds $\text{Br}^{\text{Au}}(\mu\to e)<7\times10^{-13}$, which translate
into the bound $\sqrt{|g_{L}^{\mu e}|^{2}+0.74|g_{R}^{\mu e}|^{2}}<5.48\times10^{-11}$.
We see that muon conversion can be quite sensitive to FV, which makes
it ideal for such searches.

With regards to future experiments, the Mu2e experiment~\cite{Mu2e:2014fns}
is designed to use aluminum as its target material, with a projected
sensitivity of $\text{Br}^{\text{Al}}(\mu\to e)<10^{-16}$~\cite{Kargiantoulakis:2019rjm}.
This can yield the bound $\sqrt{|g_{L}^{\mu e}|^{2}+0.74|g_{R}^{\mu e}|^{2}}<9.49\times10^{-13}$.

\subsection{Muonium-antimuonium Oscillations}

Muonium is a bound state of an electron and an antimuon, which can
oscillate to antimuonium, a bound state of a muon and a positron.
In the SM, muonium-antimuonium oscillation is heavily suppressed ($\mathcal{O}(10^{-30})$),
however, the transition probability can be sizable if there is a BSM
contribution. The time-integrated $M \to \overline{M}$
oscillation is constrained by the MACS experiment at PSI to be $P(M \to \overline{M})<8.3\times10^{-11}$~\cite{Willmann:1998gd}.
This can be used to set bounds on the FV couplings of the $Z$ to
$\mu e$.

Muonium-antimuonium oscillation through FV $Z$ couplings can proceed
via the $s$- and $t$- channels as shown in Figure~\ref{fig9}. The
details of how to calculate the time-integrated transition probability
is provided in Appendix~\ref{appendix_d}, with the transition probability
provided in Eq.~(\ref{eq:Transition_p}). Thus, the bound is obtained
by requiring that $P_{\text{Th}}<P_{\text{MACS}}$. Given that the
MACS experiment used a magnetic field of strength $B=0.1$ Tesla,
we obtain the bound 
\begin{equation}
(|g_{L}^{\mu e}|^{2}+|g_{R}^{\mu e}|^{2})^{2}+6.5|g_{L}^{\mu e}|^{2}|g_{R}^{\mu e}|^{2}-2.32(|g_{L}^{\mu e}|^{2}+|g_{R}^{\mu e}|^{2})\text{Re}(g_{L}^{\mu e}g_{R}^{\mu e*})<2.72\times10^{-6}.\label{eq:mu_osc_bound}
\end{equation}

\begin{figure}[t!]
\begin{centering}
\includegraphics[width=0.7\textwidth]{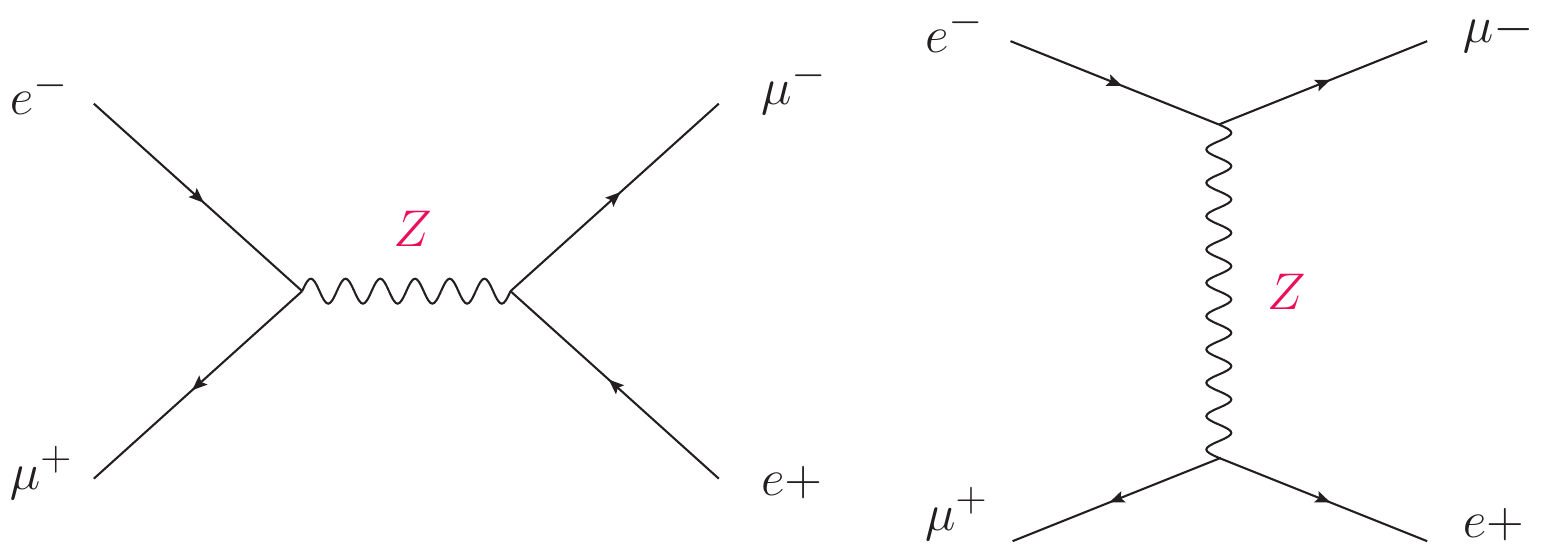} 
\par\end{centering}
\caption{Muonium-antimuonium oscillation through FV $Z$ couplings.}
\label{fig9} 
\end{figure}

\subsection{The $g-2$ Anomaly}

Theoretically, it is possible for FV $Z$ couplings to solve the $g-2$
anomaly via a FV $Z$ loop correction to the $\gamma\mu\mu$ vertex,
however, given that the size of the couplings needed to solve the
anomaly is $g\gtrsim\mathcal{O}(10^{-3})$ as suggested by the proposed
solution with a $Z'$, we can see that the above bounds exclude this
possibility.\\

\begin{figure}[ht]
\includegraphics[width=0.49\textwidth]{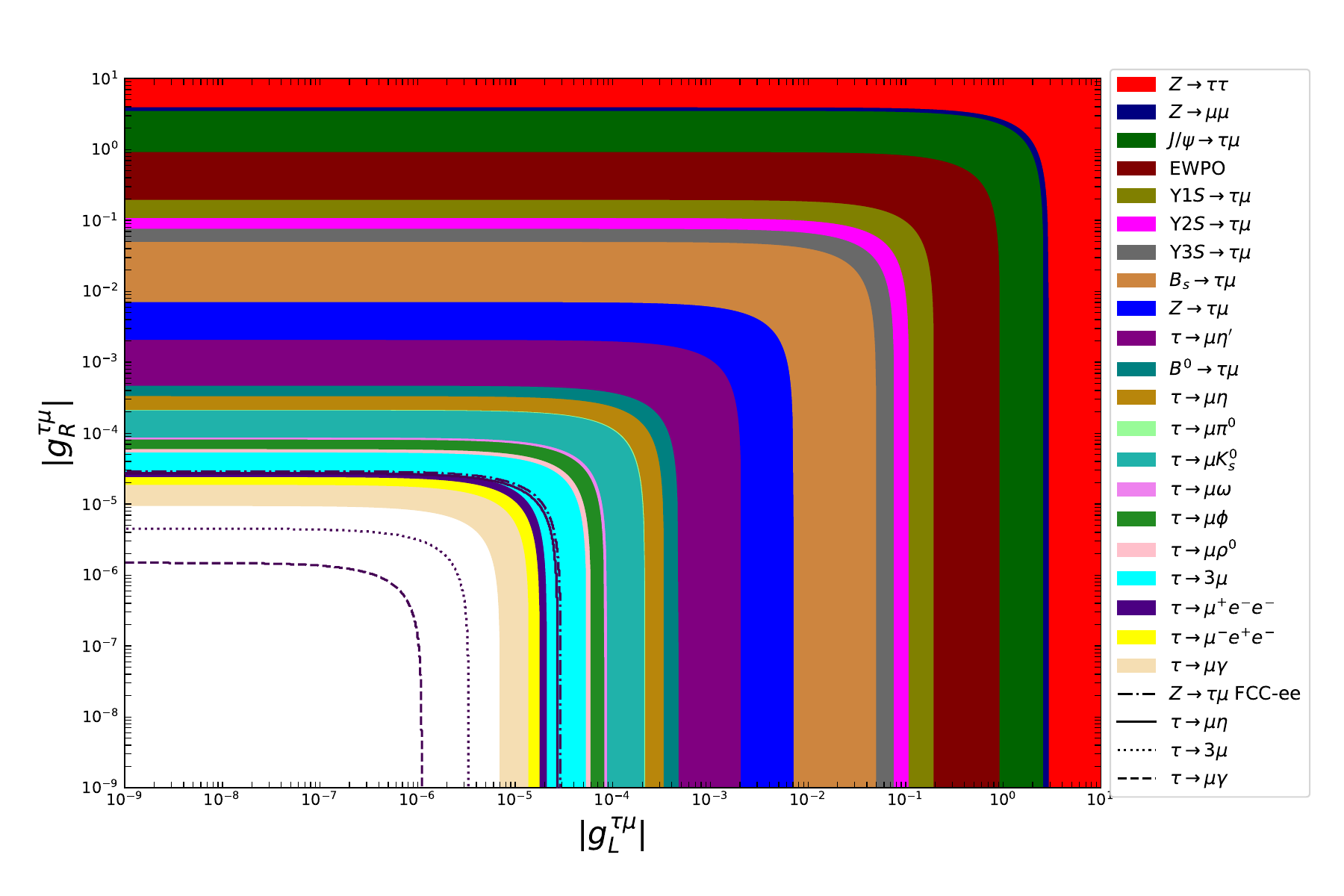}~\includegraphics[width=0.49\textwidth]{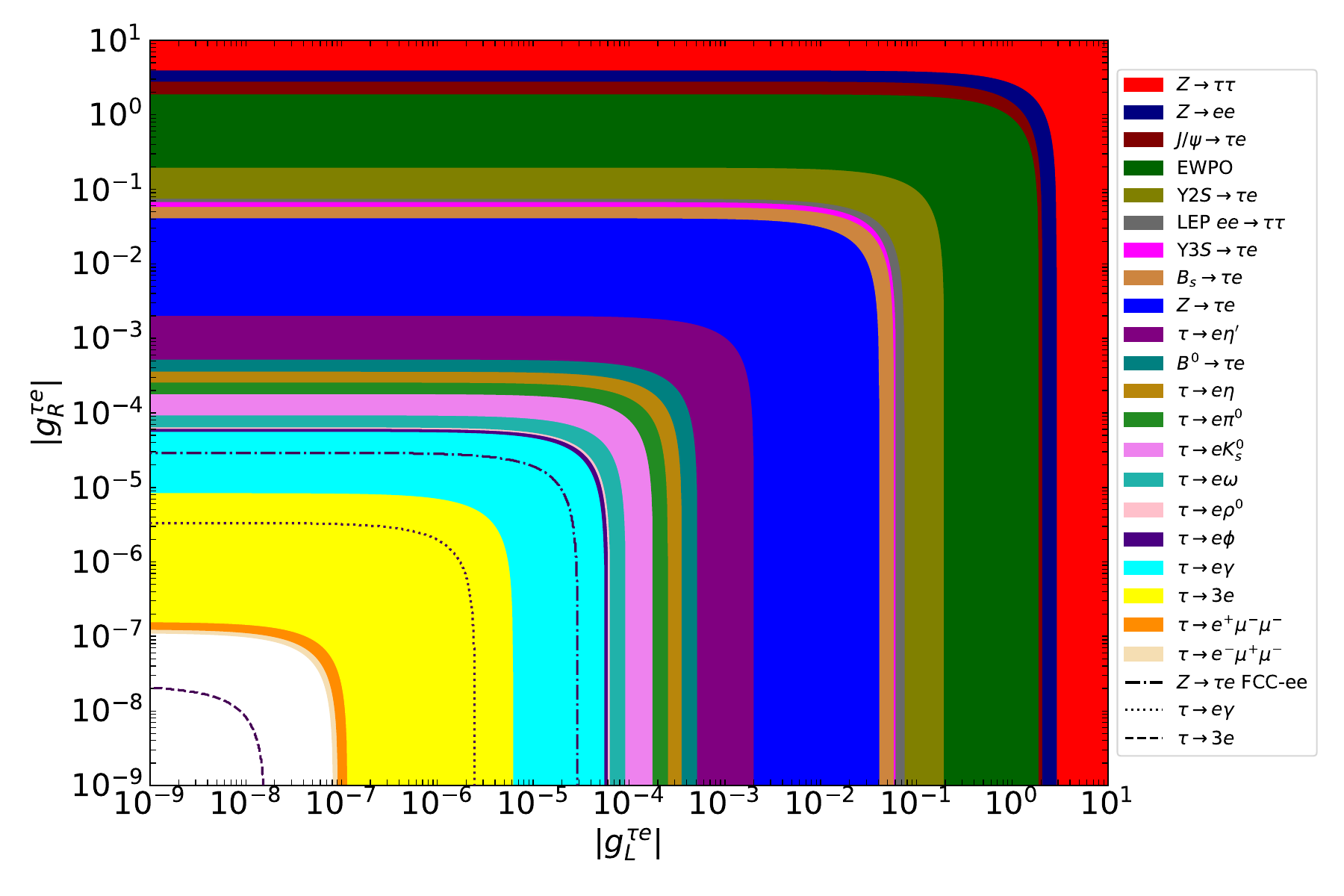}
\includegraphics[width=0.49\textwidth]{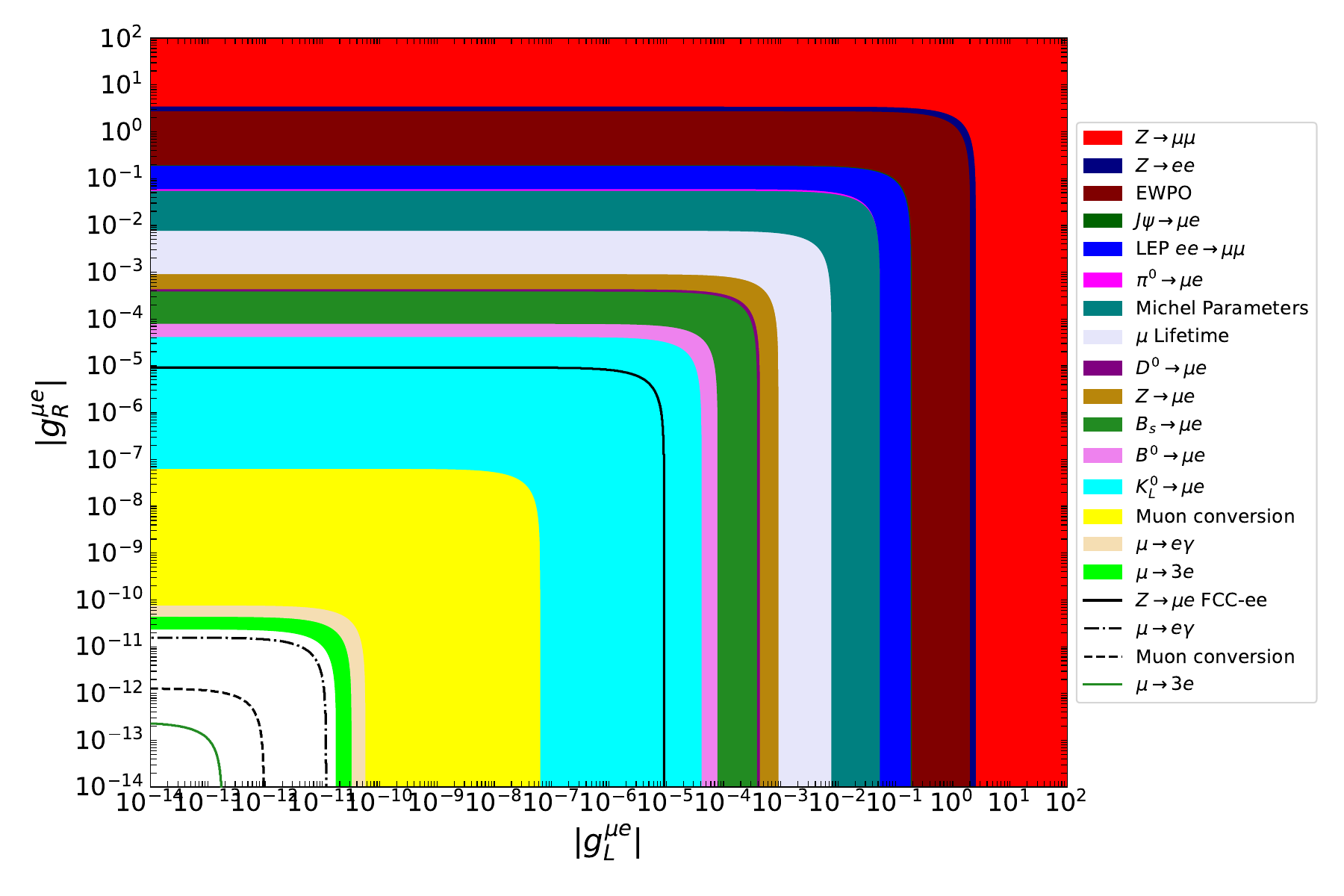} \caption{{\small{}Summary of all bounds and projections on the FV $Z$ couplings
to $\tau\mu$ (top left), $\tau e$ (top right), and $\mu e$ (bottom).}}
\label{fig10} 
\end{figure}

We conclude this section by summarizing all the bounds and projections
in Figure~\ref{fig10}.

\section{Conclusions and Outlook}

\label{sec:4} In this paper, we have studied and set upper bounds
on the FV couplings of the $Z$ boson to charged leptons. Specifically,
we have studied the bounds from direct searches from the LHC, LEP
$e^{+}e^{-}\to\mu^{+}\mu^{-},\tau^{+}\tau^{-}$ searches, FV loop
corrections to flavor-conserving $Z$ decays to leptons, corrections
to the EWPO, $\ell_{i}\to\ell_{j}\gamma$ decays, $\ell_{i}\to3\ell_{j}$
decays, corrections to the Michel parameters and to $G_{F}$, FV leptonic
meson decays, $\tau\to\mu(e)+\text{meson}$ decays, muon conversion
in nuclei and from muonium-antimuonium oscillation.

For $Z$ couplings to $\tau\mu$, we found that $\tau\to\mu\gamma$
yielded the strongest bounds, at a level of $\mathcal{O}(10^{-5})$,
followed by $\tau\to3\mu$. We also found that future experiments
could improve the bound to $\mathcal{O}(10^{-6})$. On the other hand,
for $Z$ couplings to $\tau e$, we found that $\tau\to3e,\mu\mu e$
yielded the strongest bounds, reaching a level of $\mathcal{O}(10^{-7})$
as well, followed by $\tau\to e+\gamma$. We also found that future
experiments involving $\tau\to3e$ can improve the bound to $\mathcal{O}(10^{-8})$.
Finally, for $Z$ couplings to $\mu e$, we found that $\mu\to3e$
yielded the strongest bounds, reaching a level of $\mathcal{O}(10^{-11})$,
followed by $\mu\to e\gamma$, muon conversion in nuclei, $K_{L}^{0}\to \mu e$
and $\mu\to e+\text{inv}$. We also found that future experiments
involving $\mu\to3e$ can improve the bound to $\mathcal{O}(10^{-13})$.
In addition, we also found that bounds obtained from meson decays
and from $\tau$ decays involving mesons, could yield significant
bounds, ranging between $\mathcal{O}(10^{-2})-\mathcal{O}(10^{-8})$,
thereby representing promising venues for searching for FV in the
$Z$ sector, in addition to the processes highlighted above. In general,
we find that bounds on the $Z$ couplings to $\mu e$ are several
orders of magnitude stronger than those to $\tau\mu$ and $\tau e$.

In this study, we limited ourselves to FV in the $Z$ interactions
with leptons. FV $Z$ couplings to quarks were treated in~\cite{Abu-Ajamieh:2025jov}.
Furthermore, and as mentioned in the introduction, there has been
significant interest in studying the possibility of FV in the Higgs
sector, with stringent bounds already existing in the literature (see
for instance~\cite{Harnik:2012pb}). Some of the processes discussed
herein were also used to set bounds on the FV couplings of the Higgs,
however, as several of the processes discussed here are novel, such
as the decays involving mesons, it would interesting to evaluate the
bounds that can be extracted from them on the Higgs interactions.
We plan to do this in a future work. We conclude by pointing out that
FV represents an interesting venue for potential BSM physics and thus
merits attention.

\section*{Acknowledgment}

We thank Sacha Davidson for the valuable discussion. A.A. is funded
by the university of Sharjah via the HEP RG operational grant. The
work of N.O. is supported in part by the United States Department
of Energy Grant, No.~DE-SC0012447.

\appendix

\section{Detailed Calculation of the $\ell_{i}\to\ell_{j}\gamma$ Decay~\label{appendix_a}}

Here, we provide the detailed derivation of the decay width of $\ell_{i}\to\ell_{j}\gamma$.
In our calculation, we will drop all terms $\mathcal{O}(\frac{m_{i,j}^{2}}{m_{Z}^{2}})$.
At this level of approximation, the leg corrections where the photon
is radiated from outside the loop are subleading and can be dropped.
The matrix element of the Feynman diagrams shown in Figure~\ref{fig5}
is given by 
\begin{align}
\mathcal{M} & =e\bar{u}(q_{2})\Big[\gamma^{\mu}(g_{L}^{ij*}P_{R}+g_{R}^{ij*}P_{L})I(p,q_{1},m_{i})(g_{L}P_{L}+g_{R}P_{R})\nonumber \\
 & +\gamma^{\mu}(g_{L}P_{R}+g_{R}P_{L})I(p,q_{1},m_{j})(g_{L}^{ij*}P_{L}+g_{R}^{ij*}P_{R})\Big]u(p)\epsilon_{\mu}^{*}(q_{1}),
\end{align}
where the loop integral is given by 
\begin{equation}
I(p,q,m)=\int\frac{d^{4}k}{(2\pi)^{4}}\frac{\gamma^{\beta}(\slashed{p}-\slashed{k}-\slashed{q}_{1}+m)\gamma^{\nu}(\slashed{p}-\slashed{k}+m)\gamma^{\alpha}(g_{\alpha\beta}-k_{\alpha}k_{\beta}/m_{Z}^{2})}{[(p-k-q)^{2}-m^{2}][(p-k)^{2}-m^{2}][k^{2}-m_{Z}^{2}]}.\label{eq:l_gamma_int}
\end{equation}
Evaluating the integral using dimensional regularization in the $\overline{\text{MS}}$
scheme, and dropping $m_{j}$ since $m_{\tau}\gg m_{\mu}\gg m_{e}$,
we obtain 
\begin{equation}
\mathcal{M}\simeq\frac{3ie}{16\pi^{2}}\bar{u}(q_{2})\gamma^{\mu}(g_{L}g_{L}^{ij*}P_{L}+g_{R}g_{R}^{ij*}P_{R})u(p)\epsilon_{\mu}^{*}(q_{1}),\label{eq:l_gamma_mat_2}
\end{equation}
and upon integrating $|\mathcal{M}|^{2}$ over the phase space, we
obtain Eq.~(\ref{eq:l_gamma_width}).

\section{Detailed Calculation of $\ell_{i}\to\ell_{j}+\ell_{k}+\bar{\ell}_{k}$~\label{appendix_b}}

We have the process $\ell_{i}(p)\to\ell_{j}(k_{1})+\ell_{k}(k_{2})+\bar{\ell}_{k}(k_{3})$
with $p=k_{1}+k_{2}+k_{3}$, occurring via the diagrams shown in Figure~\ref{fig6}.
The amplitudes mediated by the gauge boson $Z$ and the photon are
given by 
\begin{align}
\mathcal{M}_{Z}=\frac{-i(g_{\mu\nu}-\frac{(k_{2}+k_{3})_{\mu}(k_{2}+k_{3})_{\nu}}{m_{Z}^{2}})}{(k_{2}+k_{3})^{2}-m_{Z}^{2}}\bar{u}(k_{2})\gamma^{\nu}(g_{L}P_{L}+g_{R}P_{R})v(k_{3})\bar{u}(k_{1})\gamma^{\mu}(g_{L}^{ij*}P_{L}+g_{R}^{ij*}P_{R})u(p).
\end{align}
and 
\begin{align}
\mathcal{M}_{\gamma}=\frac{3ie}{16\pi^{2}}\frac{-ig_{\mu\nu}}{(k_{2}+k_{3})^{2}}\bar{u}(k_{2})(-ie\gamma^{\nu})v(k_{3})\bar{u}(k_{1})\gamma^{\mu}(g_{L}^{ij*}P_{L}+g_{R}^{ij*}P_{R})u(p).
\end{align}

Here, the photon contribution is the dominant one as the Z contribution
is suppressed by 5 orders of magnitude. Using the standard techniques,
the averaged squared amplitude is obtained and the differential decay
width is simplified as~\cite{Dalitz}
\begin{align}
d\Gamma & =\frac{1}{2m_{i}(2\pi)^{5}}\frac{d^{3}k_{1}}{2E_{1}}\frac{d^{3}k_{2}}{2E_{2}}\frac{d^{3}k_{3}}{2E_{3}}\delta^{(4)}(p-k_{1}-k_{2}-k_{3})|\mathcal{\bar{M}}|^{2}=\frac{1}{256\pi^{3}m_{i}^{3}}|\mathcal{\bar{M}}|^{2}dm_{12}^{2}dm_{23}^{2},
\end{align}
where the invariant quantities $m_{ij}^{2}=(k_{i}+k_{j})^{2}$ are
related by the identity $m_{23}^{2}+m_{12}^{2}+m_{13}^{2}=m_{i}^{2}+m_{j}^{2}+2m_{k}^{2}$.
After the intergation, the decay width $\Gamma(\ell_{i}\to\ell_{j}+\ell_{k}+\bar{\ell}_{k})$
can be written as 
\begin{equation}
\Gamma(\ell_{i}\to\ell_{j}+\ell_{k}+\bar{\ell}_{k})=\frac{9\alpha^{2}m_{i}}{2048\pi^{5}}\left\{ [g_{L}^{2}|g_{L}^{ij}|^{2}+g_{R}^{2}|g_{R}^{ij}|^{2}]I_{j,k}^{(1)}+g_{L}g_{R}[\Re(g_{L}^{ij})\Re(g_{R}^{ij})+\Im(g_{L}^{ij})\Im(g_{R}^{ij})]I_{j,k}^{(2)}\right\} ,
\end{equation}
with the dimensionless integrals 
\begin{align}
I_{j,k}^{(1)} & =\int_{x_{min}}^{x_{max}}dx\int_{y_{min}}^{y_{max}}dy\left\{ \frac{r_{j}(3r_{k}-2x-y+1)-r_{k}^{2}-4r_{k}x+r_{k}+2x^{2}+2xy-2x+y^{2}-y}{(r_{j}-r_{k}+y)^{2}}\right\} ,\nonumber \\
I_{j,k}^{(2)} & =\int_{x_{min}}^{x_{max}}dx\int_{y_{min}}^{y_{max}}dy\left\{ \frac{-\sqrt{r_{j}}(r_{j}-3r_{k}-y)}{(r_{j}-r_{k}+y)^{2}}\right\} ,\label{eq:int}
\end{align}
and 
\begin{align}
y_{min,max} & =\frac{1}{2x}\Big\{ r_{j}^{2}(r_{k}^{2}+x-1)-r_{k}^{4}+2r_{k}^{2}x+r_{k}^{2}-x^{2}+x\nonumber \\
 & \mp\sqrt{r_{k}^{4}-2r_{k}^{2}(x+1)+(x-1)^{2}}\sqrt{r_{j}^{4}-2r_{j}^{2}(r_{k}^{2}+x)+(r_{k}^{2}-x)^{2}}\Big\},\\
x_{min} & =(r_{j}+r_{k})^{2},\,\,\,\,\,x_{max}=(1-r_{j})^{2},\,\,r_{j}=m_{j}/m_{i},\,\,r_{k}=m_{k}/m_{i}.
\end{align}

The numerical values of the integrals in Eq.~(\ref{eq:int}) are given in Table.~\ref{Tab:int}.
by

\begin{table}[h]
\centering
\begin{tabular}{|c|c|c|}
\cline{2-3} \cline{3-3} 
\multicolumn{1}{c||}{} & $I_{j,k}^{(1)}$ & $I_{j,k}^{(2)}$\tabularnewline
\hline 
\hline 
$i=\tau,j=\mu,k=e$ & 12.714 & -0.160\tabularnewline
\hline 
$i=\tau,j=\mu,k=\mu$ & 60.188 & -5.543\tabularnewline
\hline 
$i=\tau,j=e,k=e$ & 4397.277 & -1.952\tabularnewline
\hline 
$i=\mu,j=e,k=e$ & 1510.299 & -6.172\tabularnewline
\hline 
\end{tabular}
\caption{Numerical values of the integrals in Eq.~(\ref{eq:int}).}
\label{Tab:int}
\end{table}

\section{Detailed Calculation of the Muon Conversion in Nuclei~\label{appendix_c}}

Following the convention of~\cite{Kitano:2002mt}, we can write the
most general effective Lagrangian for muon conversion in nuclei as
follows 
\begin{align}
\mathcal{L} & =c_{L}\frac{em_{\mu}}{8\pi^{2}}(\bar{e}\sigma^{\alpha\beta}P_{L}\mu)F_{\alpha\beta}-\frac{1}{2}\sum_{q}\Big[g_{LS}^{q}(\bar{e}P_{R}\mu)(\bar{q}q)+g_{LP}^{q}(\bar{e}P_{R}\mu)(\bar{q}\gamma_{5}q)+g_{LV}^{q}(\bar{e}\gamma^{\alpha}P_{L}\mu)\nonumber \\
 & \times(\bar{q}\gamma_{\alpha}q)+g_{LA}^{q}(\bar{e}\gamma^{\alpha}P_{L}\mu)(\bar{q}\gamma_{\alpha}\gamma_{5}q)+\frac{1}{2}g_{LT}^{q}(\bar{e}\sigma^{\alpha\beta}P_{R}\mu)(\bar{q}\sigma_{\alpha\beta}q)+(L\to  R)\Big],
\end{align}
where the first term arises from the magnetic dipole operator after
integrating out the loop, the first term inside the square brackets
is the scalar operator, the second is the pseudoscalar operator, the
third and fourth are the vector and axial vector operators respectively,
and the last one is the tensor operator. Inspecting the Feynman diagrams
in Figure~\ref{fig8}, we can see that we don't have any diagrams contributing
to the scalar, pseudoscalar, or tensor operators. Thus we have $g_{LS}^{q}=g_{RS}^{q}=g_{LP}^{q}=g_{RP}^{q}=g_{LT}^{q}=g_{RT}^{q}=0$.
On the other hand, the first diagram contributes to $g_{LV}^{q},g_{RV}^{q},g_{LA}^{q}$
and $g_{RA}^{q}$, whereas the second and the third diagrams contribute
to all four remaining operators. Now, to extract coefficients $c_{L,R}$,
we can use the magnetic dipole operator to write the effective vertex
after integrating out the loops in the second and third diagrams in
Figure ~\ref{fig8} as follows~ 
\begin{equation}
\mathcal{M}=\frac{em_{\mu}}{4\pi^{2}}\bar{u}(q_{2})\Big[c_{L}\slashed{q}_{1}\gamma^{\nu}P_{L}+c_{R}\slashed{q}_{1}\gamma^{\nu}P_{R}+c_{L}^{*}P_{R}\gamma^{\nu}\slashed{q}_{1}+c_{R}^{*}P_{L}\gamma^{\nu}\slashed{q}_{1}\Big]u(p)\epsilon^{*}(q_{1}),\label{eq:eff_mat}
\end{equation}
which is to be matched to the exact result found in Eq.~(\ref{eq:l_gamma_mat_2}).
A simple calculation yields the matching 
\begin{align}
c_{L}= & -\frac{3}{8m_{\mu}^{2}}g_{R}g_{R}^{\mu e*},\\
c_{R}= & -\frac{3}{8m_{\mu}^{2}}g_{L}g_{L}^{\mu e*}.
\end{align}
Turning our attention to extracting the vector and axial coefficients,
first notice that the matrix element of the first diagram is given
by 
\begin{equation}
\mathcal{M}_{1}=\bar{u}(k_{1})\gamma^{\alpha}(g_{L}^{\mu e*}P_{L}+g_{R}^{\mu e*}P_{R})u(p_{1})\frac{\Big[g_{\alpha\beta}-\frac{1}{m_{Z}^{2}}(p_{1}-k_{1})_{\alpha}(p_{1}-k_{1})_{\beta}\Big]}{(p_{1}-k_{1})^{2}-m_{Z}^{2}}\bar{u}(k_{2})\gamma^{\beta}(g_{V}^{q}-g_{A}^{q}\gamma_{5})u(p_{2}),\label{eq:VA_matrix_1}
\end{equation}
where $g_{V}^{q}$ and $g_{A}^{q}$ are the vector and axial couplings
of the $Z$ to quarks. In the non-relativistic limit, we can assume
that $E_{\mu}\sim m_{\mu}$, $E_{N}\sim m_{N}$ and that the momentum
transfer to the nucleus is negligible, in which case, Eq.~(\ref{eq:VA_matrix_1})
can be approximated as 
\begin{equation}
\mathcal{M}_{1}=-\frac{1}{m_{Z}^{2}}\times\bar{u}(k_{1})\gamma^{\alpha}(g_{L}^{\mu e*}P_{L}+g_{R}^{\mu e*}P_{R})u(p_{1})\bar{u}(k_{2})\gamma_{\alpha}(g_{V}^{q}-g_{A}^{q}\gamma_{5})u(p_{2}).\label{eq:VA_matrix_2}
\end{equation}
On the other hand, the contribution of the second and third diagrams
of Figure (\ref{fig8}) can be obtained using Eq.~(\ref{eq:l_gamma_mat_2}),
which after some simplification reads 
\begin{equation}
\mathcal{M}_{2}\simeq-\frac{3e^{2}}{32\pi^{2}}\frac{1}{(p_{1}-k_{1})^{2}}\bar{u}(k_{1})\gamma^{\alpha}(g_{L}g_{L}^{\mu e*}P_{L}+g_{R}g_{R}^{\mu e*}P_{R})u(p_{1})\bar{u}(k_{2})\gamma_{\alpha}u(p_{2}),\label{eq:VA_matrix_3}
\end{equation}
and the total contribution is given by $\mathcal{M}=\mathcal{M}_{1}+\mathcal{M}_{2}$.
Now, it's easy to extract the vector and axial coefficients, which
are found to be 
\begin{align}
g_{LV}^{q} & \simeq\Big(\frac{2}{m_{Z}^{2}}g_{V}^{q}+\frac{3\alpha}{4\pi m_{\mu}^{2}}g_{L}\Big)g_{L}^{\mu e*},\nonumber \\
g_{RV}^{q} & \simeq\Big(\frac{2}{m_{Z}^{2}}g_{V}^{q}+\frac{3\alpha}{4\pi m_{\mu}^{2}}g_{R}\Big)g_{R}^{\mu e*},\nonumber \\
g_{LA}^{q} & =-\frac{2}{m_{Z}^{2}}g_{A}^{q}g_{L}^{\mu e*},\nonumber \\
g_{RA}^{q} & =-\frac{2}{m_{Z}^{2}}g_{A}^{q}g_{R}^{\mu e*}.
\end{align}

Now that we have all the EFT coefficients, we can proceed with calculating
the $\mu\to e$ conversion rate. Following~\cite{Kitano:2002mt},
the matrix elements that contribute to the conversion are $\bra{N}\bar{q}q\ket{N}$,
$\bra{N}\bar{q}\gamma^{\mu}q\ket{N}$ and $\bra{N}\bar{q}F^{\mu\nu}q\ket{N}$,
whereas $\bra{N}\bar{q}\gamma^{\mu}\gamma_{5}q\ket{N}$ vanishes.
This implies that we can neglect $g_{LR,A}^{q}$. Therefore, we can
express the conversion rate as 
\begin{equation}
\Gamma(\mu\to e)=\Big|-\frac{e}{16\pi^{2}}c_{R}D+\Tilde{g}_{LV}^{(p)}V^{(p)}+\Tilde{g}_{LV}^{(n)}V^{(n)}\Big|^{2}+\Big|-\frac{e}{16\pi^{2}}c_{L}D+\Tilde{g}_{RV}^{(p)}V^{(p)}+\Tilde{g}_{RV}^{(n)}V^{(n)}\Big|^{2},\label{eq:mu_e_conV_rate}
\end{equation}
where 
\begin{align}
\Tilde{g}_{LV,RV}^{(p)} & =2g_{LV,RV}^{(u)}+g_{LV,RV}^{(d)},\nonumber \\
\Tilde{g}_{LV,RV}^{(n)} & =g_{LV,RV}^{(u)}+2g_{LV,RV}^{(d)},
\end{align}
and the coefficients $D$, $V^{(p)}$ and $V^{(n)}$ are the overlap
integrals between the muon, electron and nucleus wavefunctions, which
can be found for various target materials in~\cite{Kitano:2002mt}.
Notice that for neutrons, the second term of $g_{LV,RV}^{(n)}$ vanishes
due to the lack of an electric charge. According to the SINDRUM II
collaboration~\cite{SINDRUMII:2006dvw}, gold yields the strongest
bound on the conversion rate 
\begin{equation}
\text{Br}^{\text{Au}}(\mu\to e)=\Bigg[\frac{\Gamma(\mu\to e)}{\Gamma_{\text{Capture}}(\mu)}\Bigg]_{\text{Au}}<7\times10^{-13}\hspace{2mm}@\hspace{1mm}90\%\hspace{2mm}\text{CL},\label{eq:mu_e_con_bound}
\end{equation}
where the muon capture rate in gold is given by $\Gamma_{\text{Capture}}^{\text{Au}}=13.07\times10^{6}$
$\text{s}^{-1}$, and the overlap integrals for gold are given by
$V_{\text{Au}}^{(p)}=0.0974m_{\mu}^{5/2}$, $V_{\text{Au}}^{(n)}=0.146m_{\mu}^{5/2}$,
and $D_{\text{Au}}=0.189m_{\mu}^{5/2}$. On the other hand, the Mu2e
experiment~\cite{Mu2e:2014fns} is designed to use aluminum as a
target material, and is projected to have a sensitivity of~\cite{Kargiantoulakis:2019rjm}
\begin{equation}
\text{Br}^{\text{Al}}(\mu\to e)=\Bigg[\frac{\Gamma(\mu\to e)}{\Gamma_{\text{Capture}}(\mu)}\Bigg]_{\text{Al}}<10^{-16}\hspace{2mm}@\hspace{1mm}90\%\hspace{2mm}\text{CL},\label{eq:mu_e_con_proj}
\end{equation}
and for aluminum, the muon capture rate is given by $\Gamma_{\text{Capture}}^{\text{Al}}=0.7054\times10^{6}$
$\text{s}^{-1}$, and the overlap integrals for aluminum are given
by $V_{\text{Al}}^{(p)}=0.0161m_{\mu}^{5/2}$, $V_{\text{Al}}^{(n)}=0.0173m_{\mu}^{5/2}$,
and $D_{\text{Al}}=0.0362m_{\mu}^{5/2}$.

\section{Detailed Calculation of $M \to \overline{M}$ Oscillation~\label{appendix_d}}

We can write the interaction Lagrangian of the muonium-antimuonium
oscillation as follows~\cite{Conlin:2020veq} (see also~\cite{Clark:2003tv})
\begin{equation}
\mathcal{L}_{M\to \overline{M}}=-\sum_{i=1}^{5}\frac{G_{i}}{\sqrt{2}}Q_{i},\label{eq:Mu_osc_Lag}
\end{equation}
where $Q_{i}$ are all the possible operators that can induce $M\to \overline{M}$
oscillation, which are given by 
\begin{align}
Q_{1} & =[\bar{\mu}\gamma_{\alpha}(1-\gamma_{5})e][\bar{\mu}\gamma^{\alpha}(1-\gamma_{5})e],\\
Q_{2} & =[\bar{\mu}\gamma_{\alpha}(1+\gamma_{5})e][\bar{\mu}\gamma^{\alpha}(1+\gamma_{5})e],\\
Q_{3} & =[\bar{\mu}\gamma_{\alpha}(1+\gamma_{5})e][\bar{\mu}\gamma^{\alpha}(1-\gamma_{5})e],\\
Q_{4} & =[\bar{\mu}(1-\gamma_{5})e][\bar{\mu}(1-\gamma_{5})e],\\
Q_{5} & =[\bar{\mu}(1+\gamma_{5})e][\bar{\mu}(1+\gamma_{5})e],
\end{align}
with all other operators related to the 5 above by linear combination
and Fierz identities. Inspecting the Feynman diagrams in 	 ~\ref{fig9},
it is quite obvious that only $Q_{1}$, $Q_{2}$ and $Q_{3}$ contribute
to the oscillation. In~\cite{Clark:2003tv}, the time-integrated
transition probability was calculated, however, they did not take
into account the magnetic-field dependence, which was treated in~\cite{Fukuyama:2021iyw}.
The treatment of~\cite{Fukuyama:2021iyw} was applied in~\cite{Kriewald:2022erk}
to the $Z'$, so we use their results. The theoretical time-integrated
transition probability was found to be 
\begin{align}
P_{\text{Th}} & =\frac{64}{\pi^{2}}m_{\text{red}}^{6}\alpha^{6}\tau^{2}G_{F}^{2}\Big[|C_{0,0}|^{2}\Big|-G_{3}+\frac{G_{1}+G_{2}-\frac{1}{2}G_{3}}{\sqrt{1+X^{2}}}\Big|^{2}+|C_{1,0}|^{2}\Big|G_{3}+\frac{G_{1}+G_{2}-\frac{1}{2}G_{3}}{\sqrt{1+X^{2}}}\Big|^{2}\Big],\nonumber \\
 & =\frac{2.57\times10^{-5}}{G_{F}^{2}}\Big[|C_{0,0}|^{2}\Big|-G_{3}+\frac{G_{1}+G_{2}-\frac{1}{2}G_{3}}{\sqrt{1+X^{2}}}\Big|^{2}+|C_{1,0}|^{2}\Big|G_{3}+\frac{G_{1}+G_{2}-\frac{1}{2}G_{3}}{\sqrt{1+X^{2}}}\Big|^{2}\Big],
\label{eq:Transition_p}
\end{align}
where $G_{F}$ is the Fermi constant, $\tau_{\mu}=2.2~\mu s$ is the
lifetime of the muon, $m_{\text{red}}=m_{\mu}m_{e}/(m_{\mu}+m_{e})\simeq m_{e}$
is the reduced mass of the muonium, $X$ encodes the magnetic flux
density, which is given by $X\simeq6.31B/\text{Tesla}$ (see the Appendix
B of~\cite{Kriewald:2022erk} for more detail), and $|C_{F,m}|^{2}$
denote the population of the muonium states with a total angular momentum
$=F$ and an angular momentum in the $\hat{z}$ direction $=m$. In
our calculation, we use the same values quoted in~\cite{Fukuyama:2021iyw},
which are $|C_{0,0}|^{2}=0.32$ and $|C_{1,0}|^{2}=0.18$.

Now, the coefficients $G_{i}$ are model-dependent, and have been
found for $Z'$ in~\cite{Kriewald:2022erk}. Thus, we can easily
extend their results to the case of a FV $Z$. We find 
\begin{align}
G_{1} & =\frac{\sqrt{2}|g_{L}^{\mu e}|^{2}}{8m_{Z}^{2}},\\
G_{2} & =\frac{\sqrt{2}|g_{R}^{\mu e}|^{2}}{8m_{Z}^{2}},\\
G_{3} & =\frac{2\sqrt{2}g_{L}^{\mu e}g_{R}^{\mu e*}}{8m_{Z}^{2}}.
\end{align}

\end{document}